\documentclass[11pt,a4paper]{article}
\pdfoutput=1
 
\usepackage{amsmath,amsthm,amssymb}
\usepackage{graphics,graphicx}
\usepackage{epsfig}
\usepackage{multicol}
\usepackage{color}
\makeatletter
\@addtoreset{equation}{section}
\makeatother
\renewcommand{\theequation}{\thesection.\arabic{equation}}

\usepackage{braket}

\begin{document}

\begin{flushright}
%{RESCEU-37/12}
\end{flushright}
\begin{center}
\LARGE{\bf The Compton$-$Schwarzschild correspondence from extended de Broglie relations}
\end{center}

\begin{center}
\large{\bf Matthew J. Lake, } ${}^{a,b}$\footnote{matthewj@nu.ac.th} \large{\bf and Bernard Carr} ${}^{c}$\footnote{b.j.carr@qmul.ac.uk} 
\end{center}
\begin{center}
\emph{ ${}^a$ The Institute for Fundamental Study, ``The Tah Poe Academia Institute", \\ Naresuan University, Phitsanulok 65000, Thailand \\}
\emph{ ${}^b$ Thailand Center of Excellence in Physics, Ministry of Education, Bangkok 10400, Thailand \\}
\emph{ ${}^c$ School of Physics and Astronomy, Queen Mary University of London, Mile End Road, London E1 4NS, UK \\}
%\emph{ ${}^d$Research Center for the Early Universe (RESCEU), Graduate School of Science, \\ University of Tokyo, Tokyo 113-0033, Japan \\} 
%\emph{ ${}^e$  Department of Mathematics, University College London, Gower Street, \\ London, WC1E 6BT, United Kingdom}
\vspace{0.1cm}
\end{center}

%Abstract%%%%%%%%%%%%%%%%%%%%%%%%%%%%%%%%%%%%%%%%%%%%%%%%%%%%%%%%%%%%%%%%``
%%%%%%%%%%%%%%%%%%%%%%%%%%%%%%%%%%%%%%%%%%%%%%%%%%%%%%%%%%%%%%%%%%%%%

\begin{abstract}
The Compton wavelength gives the minimum radius within which the mass of a particle may be localized due to quantum effects, while the Schwarzschild radius gives the maximum radius within which the mass of a black hole may be localized due to classical gravity. In a mass-radius diagram, the two lines intersect near the Planck point $(l_P,m_P)$, where quantum gravity effects become significant. Since canonical (non-gravitational) quantum mechanics is based on the concept of wave-particle duality, encapsulated in the de Broglie relations, these  relations should break down near $(l_P,m_P)$. It is unclear what physical interpretation can be given to quantum particles with energy $E \gg m_Pc^2 $, since they correspond to wavelengths $\lambda \ll l_P$ or time periods $\tau \ll t_P$ in the standard theory. We therefore propose a correction to the standard de Broglie relations, which gives rise to a modified Schr{\" o}dinger equation and a modified expression for the Compton wavelength, which may be extended into the region $E \gg m_Pc^2$. For the proposed modification, we recover the expression for the Schwarzschild radius for $E \gg m_Pc^2$ and the usual Compton formula for $E \ll m_Pc^2$. 
The sign of the inequality obtained from the uncertainty principle reverses at $m \approx m_P$, so that the Compton wavelength and event horizon size may be interpreted as minimum and maximum radii, respectively. We interpret the additional terms in the modified de Broglie relations as representing the self-gravitation of the wave packet.
\end{abstract}

%Section1%%%%%%%%%%%%%%%%%%%%%%%%%%%%%%%%%%%%%%%%%%%%%%%%%%%%%%%%%%%%%%%%``
%%%%%%%%%%%%%%%%%%%%%%%%%%%%%%%%%%%%%%%%%%%%%%%%%%%%%%%%%%%%%%%%%%%%%
\section{Introduction} \label{Sec.1}
In 1924, de Broglie proposed the idea of wave-particle duality as a fundamental feature of all forms of matter and energy, introducing his famous relations
\begin{eqnarray}
\label{deBroglie}
E = \hbar\omega \, , \ \ \ \vec{p} = \hbar\vec{k} \, . 
\end{eqnarray} 
By combining these with the energy-momentum relation for a non-relativistic point particle,
\begin{eqnarray} 
\label{E=p^2/2m}
E = \frac{p^2}{2m} + V \, ,
\end{eqnarray} 
Schr{\" o}dinger obtained the equation for the quantum wave function,
\begin{eqnarray} \label{Schrodinger}
\hat{H}\psi = \left(\frac{\hat{p}^2}{2m} + V\right)\psi = \left(-\frac{\hbar^2}{2m}\vec{\nabla}^2 + V\right)\psi = i\hbar\frac{\partial \psi}{\partial t} \, ,
\end{eqnarray} 
and the development of modern quantum theory began. Setting $V=0$ gives the usual dispersion relation for a free particle $\omega = (\hbar/2m)k^2$. However, these relations are thought to break down near the Planck mass and length scales
\begin{eqnarray} \label{Planck}
m_P = \sqrt{\frac{\hbar c}{G}} \, , \ \ \ l _P = ct_P = \sqrt{\frac{\hbar G}{c^3}}.% \, ,\ \ \  \omega_P = \frac{2\pi}{t_P} \, , \ \ \  k_P =\frac{ 2\pi}{l_P} \, ,
\end{eqnarray} 
Specifically, the non-relativistic energy-momentum relation $E = p^2/(2m)$ implies that the angular frequency and wave number of the matter waves reach the Planck values
\begin{eqnarray} \label{Planck*}
\omega_P = \frac{2\pi}{t_P} \, , \ \ \  k_P =\frac{ 2\pi}{l_P} \, ,
\end{eqnarray}
for free particles with energy, momentum and mass given by $E = pc = 2\pi m_Pc^2$ and $m = \pi m_P$. Since a further increase in energy would imply a de Broglie wavelength smaller than the Planck length, it is unclear how quantum particles behave for $E > 2\pi m_Pc^2$. 

Though it may be argued that the Newtonian formula is not valid close to the Planck scales due to both relativistic and gravitational effects, the former is not necessarily true in the usual sense. In non-gravitational theories, relativistic effects are important for particles with kinetic energy higher than their rest mass energy. Thus, a particle can have Planck scale energy if it has Planck scale rest mass, even if it is moving non-relativistically. We consider the rest frame of quantum particles with rest masses extending into the regime $m \gg m_P$, for which the the standard non-relativistic analysis is expected to hold except for modifications induced by gravity, rather than relativistic velocities.

In this paper, we propose extended de Broglie relations which reduce to the standard form for $E \ll m_Pc^2$ but remain valid for $E \gg m_Pc^2$. For $V=0$, the modified relations obey a dispersion relation of the form $\Omega = (\hbar/2m)\kappa^2$, where $\Omega = \Omega(\omega)$ and $\kappa = \kappa(k)$, so that the evolution of the quantum system is no longer governed by the standard Schr{\" o}dinger equation and the canonical $\hat{p}$ and $\hat{H}$ operators also change. However, the underlying mathematical formalism of canonical quantum theory -- the Hilbert space structure and the postulates connecting this to physical observables  \cite{Rae00,Ish95} -- are not disturbed. What \emph{does} change is the equations of motion (EOM) and the corresponding expressions for the canonical operators. Specifically, the plane-wave modes that are the simultaneous eigenfunctions of the \emph{usual} $\hat{p}$ and $\hat{H}$ operators are also extended to include wavelengths corresponding to $E > 2\pi m_Pc^2$ and $p > 2\pi m_Pc$. Interestingly, we find that this modification does not imply a positional uncertainty smaller than the Planck length. 

Crucial to understanding this result is the behavior of the modified Compton wavelength {($\lambda_C$): as $E$ increases, this decreases from large values for $E \ll m_Pc^2$, reaches a minimum at $E \approx m_Pc^2$, then increases again to large values for $E \gg m_Pc^2$. Using the standard uncertainty relation for position and momentum, now written in terms of the modified commutator, to estimate the Compton radius of wave packets expanded as superpositions of the modified $\hat{p}$ eigenfunctions, we obtain $\lambda \geq \lambda_C \approx l_Pm_P/m$ (as usual) for $m \ll m_P$ but $\lambda \leq \lambda_S \approx l_Pm/m_P$ for $m \gg m_P$. Crucially, the sign of the inequality reverses at $\lambda_C \approx l_P$, $m \approx m_P$. Therefore, we obtain a unified expression for the Compton wavelength, interpreted as the \emph{minimum} spatial extent over which the mass of a particle is distributed, and the Schwarzschild radius ($\lambda_S$), interpreted as a \emph{maximum} spatial extent within which the mass of a black hole resides.  

This unified expression has the form anticipated on the basis of the Generalized Uncertainty Principle (GUP) \cite{Ca:2013}. This involves what is termed the Black Hole Uncertainty Principle (BHUP) correspondence \cite{Ca:2014,cmp}, here referred to as the Compton-Schwarzschild correspondence. In the present formulation, this is obtained by applying the standard uncertainty principle (UP) to wave functions that are consistently defined for all energy scales. The underlying theory is an extended form of canonical non-relativistic quantum mechanics which, at first sight, appears to be non-gravitational, in the sense that the results obtained do not require the introduction of a classical Newtonian gravitational potential. Nonetheless, introducing such a potential as an external field acting on the wave packet, along with a maximum speed $v_{max} = c$, gives the Schwarzschild radius from the standard non-relativistic formula for the escape velocity. In this way, our results are \emph{consistent} with classical physics at any mass scale but the existence of the event horizon may be seen as both a gravitational and a quantum effect for $m \gtrsim m_P$. 

In fact, both gravitational and relativistic effects are introduced into the extended theory in a more subtle way, in that the extended de Broglie relations now depend on both $G$ and $c$, as well as $\hbar$. However, the dependence on $G$ and $c$ is subdominant for $E \ll m_Pc^2$, whereas the dependence on $\hbar$ is subdominant for $E \gg m_Pc^2$. This implies that quantum physics naturally becomes less `quantum' for the high energy scales associated with macroscopic objects (such as large black holes) and this may be relevant to the problem of decoherence \cite{Zurek:2003zz,Anglin:1996bb}. We therefore interpret the extended de Broglie relations as representing the effect of the particle's self-gravity on the evolution of the quantum state. The simplest way to account for the effect of the gravitational field of the particle is to introduce a classical external potential $V = -Gm/r$ that confines the wave packet within an (energy-dependent) radius about $r=0$. However, while this approach may provide a good approximation in the centre-of-mass frame, it is not obvious how to extend it to include the effect of self-gravity on an otherwise free particle. 
 
Other, more subtle, approaches such as the Schr{\" o}dinger-Newton equation (see \cite{Singh:2015sua} and references therein) suffer from similar complications and require the inclusion of at least a semiclassical potential. The approach presented here has no such drawback and correctly predicts the emergence of black holes as an essentially gravitational effect, arising in an extended quantum theory. Nonetheless, it shares some similarities with the Schr{\" o}dinger-Newton equation, in that $m|\psi|^2$ may be interpreted as a `potential' mass per unit volume, which is analogous to a classical mass density $\rho$. In some sense, the new theory is based on pushing this analogy to its limit. In a classical theory, infinitesimal mass elements $\rho dV$ interact with each other gravitationally, whereas in non-gravitational quantum mechanics, there is no self-interaction in $\psi$ corresponding to interacting mass elements $m|\psi|^2dV$. Intuitively, one way to consistently introduce such an interaction might be through the modification of the de Broglie relations to include $G$ and $c$. As a guide to developing a n{\" a}ive theory, obtained by substituting extended relations into the non-relativistic energy-momentum relation for a point particle, we require the resulting dispersion relation to recover the salient features of both canonical quantum mechanics and classical gravitational physics in the appropriate asymptotic regimes. Specifically, we wish to recover the Compton and Schwarzschild expressions for systems with $E \ll m_Pc^2$ and $E \gg m_Pc^2$, respectively, using standard heuristic arguments applied to the new relations.

The structure of this paper is as follows. Sec.~\ref{Sec.2} reviews the basic mathematical formalism of canonical non-relativistic quantum theory and its relation to physical observations. We also discuss the independence/interdependence of the postulates of quantum mechanics and how their modifications may affect each other.  Sec.~\ref{Sec.3} sets out the theory based on extended de Broglie relations. Sec.~\ref{Sec.3.1} considers the asymptotic form  required to reproduce the Schwarzschild radius for $E \gg m_Pc^2$; Sec.~\ref{Sec.3.2} proposes extended relations which correctly reproduce the standard Compton and Schwarzschild formulae in the limits $E \ll m_Pc^2$ and $E \gg m_Pc^2$ and ensure continuity of all physical observables at $E \approx m_Pc^2$; Sec.~\ref{Sec.3.3} considers the implications of the extended relations for the EOM and the form of the momentum and Hamiltonian operators. Sec.~\ref{Sec.4} derives a unified Compton-Schwarzschild expression using the UP and modified commutators and compares this with the BHUP prediction. Sec.~\ref{Sec.5} considers the implications of this theory for the Hawking temperature of black holes close to the Planck mass. Sec.~\ref{Sec.6} discusses various other issues, such as the implications of the extended de Broglie relations for GUP phenomenology \cite{Tawfik:2014zca}, and the prospects for future work. 

A technical but important issue, the time evolution of self-gravitating states, is discussed separately in Appendix A. The modified Schr{\" o}dinger equation implied by the extended relations involves higher order time derivatives, which often lead to non-unitary evolution in a variety of contexts. However, the time evolution operators for the extended de Broglie theory are shown to be non-canonical but \emph{unitary}, which raises the question of how the usual problems are avoided. Two tentative answers are proposed in Appendix B, though further work is required to answer this question definitively.
%
%Section2%%%%%%%%%%%%%%%%%%%%%%%%%%%%%%%%%%%%%%%%%%%%%%%%%%%%%%%%%%%%%%%%``
%%%%%%%%%%%%%%%%%%%%%%%%%%%%%%%%%%%%%%%%%%%%%%%%%%%%%%%%%%%%%%%%%%%%%
\section{The postulates of canonical quantum mechanics} \label{Sec.2}
The book by Rae \cite{Rae00} lists five basic postulates of quantum mechanics. These define the contents of the theory in terms of wave functions, which may be expanded in terms of a discrete set of eigenfunctions of a Hermitian operator.

\begin{enumerate}
%1
\item For every dynamical system there exists a wave function, $\psi$, that is a continuous, square-integrable, single-valued function of the parameters of the system and of time, and from which all possible predictions about the physical properties of the system can be obtained.
%2
\item Every dynamical variable may be represented by a Hermitian operator whose eigenvalues represent the possible results of carrying out a measurement of the value of the dynamical variable. Immediately after such a measurement, the wave function of the system will be identical to the eigenfunction corresponding to the eigenvalue obtained as a result of the measurement.
%3
\item The operators representing the position and momentum of a particle are $\vec{r}$ and $-i\hbar\vec{\nabla}$ respectively. Operators representing other dynamical quantities bear the same functional relation to these as do the corresponding classical quantities to the classical position and momentum variables.
%4
\item When a measurement of a dynamic variable represented by the Hermitian operator $\hat{Q}$ is carried out on a system whose wave function is $\psi$, then the probability of obtaining a particular eigenvalue $q_m$ will be $|a_m|^2$, where $\psi = \sum a_n \phi_n$ and the $\phi_n$ are the eigenfunctions corresponding to the eigenvalues $q_n$.
%5
\item Between measurements, the development of the wave function with time is governed by the time-dependent Schr{\" o}dinger equation.
\end{enumerate}
Such criteria cannot account for the effect of spin or any observable with a continuous spectrum of possible values {(though these are deliberate omissions in an introductory text). The latter problem can be fixed by replacing the discrete expansion, $\psi = \sum a_n \phi_n$, in Postulate 4 with the integral expansion, $\psi(\vec{r},t) = \int a(k) \phi(\vec{r},t;k)dk$, where $k$ is a continuous real parameter, and replacing $|a_m|^2$ with $|a(k)|^2$ and $q_m$ with $q$, a continuous real eigenvalue corresponding to $k$. To include spin, we must demonstrate the equivalence of every differential operator to a matrix operator and hence of the wave function to a vector in the phase space of the theory \cite{Rae00}. This elucidates the underlying mathematical structure of quantum theory and allows us to express the basic postulates in an alternative form. 

The book by Isham \cite{Ish95} gives four postulates for quantum mechanics:

\begin{enumerate}
%1
\item The predictions of results of measurements made on an otherwise isolated system are probabilistic in nature. In situations where the maximum amount of information is available, this probabilistic information
is represented by a vector $\ket{\psi}$ in a complex Hilbert space $\mathcal{H}$ that forms the state space of the theory. In so far as it gives the most precise predictions that are possible, this vector can be regarded as the mathematical representation of the physical state of the system.
%2
\item The observables of the system are represented mathematically by self-adjoint operators that act on the Hilbert space $\mathcal{H}$. 
%3
\item If an observable quantity $A$ and a state are represented respectively by the self-adjoint operator $\hat{A}$ and the normalized vector $\ket{\psi} \in \mathcal{H}$, then the expected result
\footnote{In order to comply with the conventions of standard probability theory, the word `average' is best reserved for the average of an actual series of measurements. When referring to the `average' predicted by the mathematical formalism, it is more appropriate to use the phrase `expected result' or `expected value'.}
$\braket{A}_{\psi}$ of measuring $A$ is
\begin{eqnarray} \label{exp_val}
\braket{A}_{\psi} = \braket{\psi|\hat{A}|\psi}.
\end{eqnarray}
%4
\item  In the absence of any external influence (i.e., for  a closed system), the state vector $\ket{\psi}$ changes smoothly according to the time-dependent Schr{\" o}dinger equation
\begin{eqnarray} \label{Schrod_Ish}
i\hbar \frac{\partial \ket{\psi}}{\partial t} = \hat{H}\ket{\psi},
\end{eqnarray}
where $\hat{H}$ is a special operator known as the Hamiltonian.
\end{enumerate}
These four rules define the  general framework within which it has so far been possible to describe \emph{all} quantum mechanical systems \cite{Ish95}. Rae's postulates define the theory in terms of a wave function $\psi$ rather than a wave vector $\ket{\psi}$ and deal only with systems with discrete spectra of eigenvalues. The latter is easily corrected, so the main difference from Isham's rules is the absence of his Postulate 3.

In Rae's formulation, the canonical definition of the momentum operator, $\hat{p} = -i\hbar\vec{\nabla}$ in Postulate 3, plus the assertion that quantum mechanical operators are defined by analogy with their classical non-relativistic counterparts, is implicit in Postulate 5. Indeed, this reasoning underpins the original derivation of the Schr{\" o}dinger equation, so it is also implicit in Isham's Postulate 4. Ultimately, this comes from assuming that quantum mechanical operators representing observables are defined by analogy with their classical equivalents, together with the assumption that the usual de Broglie relations (\ref{deBroglie}) hold. If we change these
relations, but retain the analogous classical/quantum formulae, we change the dispersion relation for matter waves and hence the required EOM for the quantum system. However, there is no need to change any of Isham's  first three postulates or to interfere with the state space structure of the theory which underpins its probabilistic interpretation \cite{Ish95}.

Such an approach is adopted here. In particular, the Hilbert space structure of quantum mechanics leads to the general uncertainty relations for two arbitrary operators  $\hat{O}_1$ and $\hat{O}_2$ \cite{Ish95}:
\begin{eqnarray} \label{SUP}
\Delta_{\psi}O_1 \Delta_{\psi}O_2 &\geq& \frac{1}{2}\sqrt{|\langle \psi|[\hat{O}_1,\hat{O}_1]|\psi\rangle|^2 + |\langle \psi|[\hat{A},\hat{B}]_{+}|\psi\rangle|^2}
\geq \frac{1}{2}|\langle \psi|[\hat{O}_1,\hat{O}_1]|\psi\rangle|, 
\end{eqnarray}
where $[\hat{O}_1,\hat{O}_2]$ is the commutator of $\hat{O}_1$ and $\hat{O}_2$ and $[\hat{A},\hat{B}]_{+}$ is the anticommutator of $\hat{A} = \hat{O}_1 - \langle \hat{O}_1\rangle_{\psi}\hat{\mathbb{I}}$ and $\hat{B} = \hat{O}_2 - \langle \hat{O}_2\rangle_{\psi}\hat{\mathbb{I}}$.
This expression remains unchanged in the theory presented below. Choosing $\hat{O}_1 = \hat{x}$ and $\hat{O}_2 = \hat{p}$, the standard expression for the commutator $[\hat{x},\hat{p}]$ is modified, leading to predictions that differ from those of canonical quantum theory close to the Planck scales. This approach thus circumvents problems associated with some of the previous attempts to obtain a unified Compton-Schwarzschild  expression, using arguments based on the GUP~\cite{Tawfik:2014zca}.
\footnote{This should not be confused with the `general uncertainty principle' in the canonical theory \cite{Ish95}, though the terminology is confusing.}
Unless based on modified dispersion relations, these necessarily violate results - such as Eq. (\ref{SUP}) -  that follow from standard quantum mechanics  \cite{Tawfik:2014zca,LaCa:2015,La15}. As discussed in \cite{LaCa:2015}, it may nevertheless be possible to obtain a GUP-type expression from a theory of deformed quantum mechanics (c.f. \cite{Pe11}). Such a theory can be consistently formulated in terms of a vector space \cite{La09,Zhang:2003wv,Hirshfeld(2002)} and may incorporate quantum gravity effects if the GUP deformations are  directly related to deformations of the metric \cite{Maziashvili:2011dx,Faizal:2015kqa}. However, this is not the approach adopted here. 

It is also worth noting that, although GUP theories may lead to unified expressions for the Compton and Schwarzschild radii, the direction of the inequality still points the wrong way for the latter. Any derivation of the Compton wavelength in a non-relativistic theory is necessarily heuristic, since this is an essentially relativistic phenomenon, marking the boundary beyond which the concept of a wave packet describing a system with conserved particle number breaks down \cite{Soviet,Greiner:1990tz,AlvarezGaume:2012zz}. However, such heuristic arguments may be placed on a firmer theoretical footing by assuming appropriate cut-offs for the eigenfunction expansions of wave vectors in the non-relativistic theory \cite{LaCa:2015}. As shown in Sec. \ref{Sec.3.1}, standard n{\" a}ive arguments, together with the extended de Broglie relations proposed for $E \gg m_Pc^2$, then lead to an interpretation  of the Schwarzschild radius as a \emph{maximum} radius within which the mass of the  `particle' (i.e. black hole) must be confined. 
%
%Section3%%%%%%%%%%%%%%%%%%%%%%%%%%%%%%%%%%%%%%%%%%%%%%%%%%%%%%%%%%%%%%%%``
%%%%%%%%%%%%%%%%%%%%%%%%%%%%%%%%%%%%%%%%%%%%%%%%%%%%%%%%%%%%%%%%%%%%%
\section{Extended de Broglie relations} \label{Sec.3}
In this section, we begin by outlining the asymptotic form of the extended de Broglie relations required to reproduce the Schwarzschild radius as an extension of the Compton wavelength formula in the limit $E \gg m_Pc^2$. We will use standard (albeit n{\" a}ive) arguments for the Compton wavelength, which is derived heuristically from the non-relativistic UP. We then propose a functional form for the extended relations that correctly reproduces the standard Compton and Schwarzschild expressions for $E \ll m_Pc^2$ and $E \gg m_Pc^2$, respectively, and also ensures the continuity of all physical observables at $E \approx m_Pc^2$. Finally, we investigate how the modified dispersion relation affects the EOM for the quantum state and the expressions for the corresponding momentum and Hamiltonian operators. The standard Schr{\" o}dinger equation is also recovered in the limit $E \ll m_Pc^2$, $\lambda \gg l_p$. 
%
%Section3.1%%%%%%%%%%%%%%%%%%%%%%%%%%%%%%%%%%%%%%%%%%%%%%%%%%%%%%%%%%%%%%``
\subsection{de Broglie relations in the limit $E \gg m_Pc^2$} \label{Sec.3.1}
Let us suppose that, in the limit $E \gg m_Pc^2$, the generalized de Broglie relations (in one dimension) take the form 
\begin{eqnarray} \label{deBroglie_m>>m_P}
E = \hbar\Omega \approx \hbar\beta\frac{\omega_P^2}{\omega} \, , \ \ \ p = \hbar \kappa \approx \hbar \beta \frac{k_P^2}{k} \, ,
\end{eqnarray} 
where $\beta > 0$ is a dimensionless constant and $\omega_P$ and $k_P$ are defined by Eq.~\eqref{Planck*}. We assume that momentum eigenfunctions with $E \gg m_P c^2$ correspond to $\omega \ll \omega_P$ and $k \ll k_P$, as in the $E \ll m_P c^2$ case, although we shall see in Sec.~\ref{Sec.3.2} that this is not the only possibility. For simplicity, we will work in one dimension throughout this paper, but the extension to three (or more) dimensions is straightforward and involves no new fundamental ideas. 

Combining Eqs. (\ref{deBroglie_m>>m_P}) and (\ref{E=p^2/2m}) with $V=0$, we have
\begin{eqnarray} \label{disp_rel_all_m}
\Omega = \frac{\hbar}{2m}\kappa^2 \, 
\end{eqnarray} 
or equivalently
\begin{eqnarray} \label{disp_rel_m>>m_P*}
\omega \approx \frac{2 G m}{(2\pi)^2 \beta c}k^2 \, .
\end{eqnarray}
If the momentum operator eigenfunctions take the usual form
\begin{eqnarray} \label{p_eigenfn_m>>m_P}
\phi(k,\omega : x,t) = \exp\left[i(k x - \omega t\right)] \, ,
\end{eqnarray} 
then substituting $\phi = \psi$ 
into the Schr{\" o}dinger-type equation,
\begin{eqnarray} \label{Schrod-type}
\frac{\partial^2 \psi}{\partial x^2} = \alpha \frac{\partial \psi}{\partial t} \, ,
\end{eqnarray} 
gives
\begin{eqnarray} \label{alpha_m>>m_P}
\alpha = - \frac{ik^2}{\omega}  \approx -\frac{i(2\pi)^2 \beta c}{2 G m} \, .
\end{eqnarray} 
The approximate EOM for the quantum state with no classical external potential may then be written as
\begin{eqnarray} \label{Schrod_1D}
\hat{H}'\psi = \frac{\hat{p}'^2}{2m}\psi \approx -\frac{mc^2}{k_P^2}\frac{\partial^2 \psi}{\partial x^2} \approx \frac{i\beta\hbar}{2} \frac{\partial \psi}{\partial t} \, .
\end{eqnarray}
In the position space representation, the differential part of the momentum operator takes the usual form but the multiplying factor differs from the canonical one. The first corresponds to the functional form of the dispersion relation, while the second sets the phenomenologically important length scale. The position operator may be defined as usual, so that 
\begin{eqnarray} \label{p_x_1D}
\hat{p}' \approx -i\frac{\sqrt{2}mc}{k_P}\frac{\partial}{\partial x} \, , \ \ \  \hat{x} = x \, .
\end{eqnarray}
The commutator of the position and momentum operators is then
\begin{eqnarray} \label{x-p_comm_m>>m_P}
[\hat{x},\hat{p}'] = i\frac{\sqrt{2}mc}{k_P} \, .
\end{eqnarray}

It is clear that the proposed asymptotic form  (\ref{deBroglie_m>>m_P}) of the extended de Broglie relations for $E \gg m_Pc^2$ only disturbs the mathematical structure of quantum mechanics in so far as the momentum operator is multiplied by a constant factor. Although this affects the EOM, as well as the definitions of operators that are functions of $\hat{p}$ and commutators involving these, it in no way affects the underlying Hilbert space structure of the theory, its probabilistic interpretation or the role of Hermitian operators. Postulates 3 and 5 of Rae \cite{Rae00} are changed but only trivially. The underlying mathematical structure and formalism is exactly that of canonical quantum mechanics, so the new theory is well defined in the $E \gg m_Pc^2$ limit. 

In the next section, we will see that, for de Broglie relations which hold for all energies but take the appropriate asymptotic forms for $E \ll m_Pc^2$  and $E \gg m_Pc^2$ (i.e. in which a Schr{\" o}dinger-type equation describes the evolution of the system), the general EOM is \emph{not} of the Schr{\" o}dinger form. It is more complicated, as are the expressions for the operators $\hat{p}$ and  $\hat{H}$. The Hilbert space structure is nonetheless preserved. Again, only Postulates 3 and 5 of Rae must be amended, this time less trivially, while the mathematical structure embodied in his Postulates 1, 2 and 4 remains the same. 

The motivation for requiring the asymptotic form (\ref{deBroglie_m>>m_P}) is that arguments analogous to those used to derive the Compton wavelength from the UP (\ref{SUP}), together with identifications suggested by the modified de Broglie relations themselves, lead to the expression for the Schwarzschild radius. From Eqs.  (\ref{SUP}) and (\ref{x-p_comm_m>>m_P}), it follows that 
\begin{eqnarray}  \label{x-p_SUP_m>>m_P}
\Delta_{\psi}x\Delta_{\psi}p' \geq \frac{1}{2}\braket{\psi| [\hat{x},\hat{p}'] |\psi} \approx \frac{m \, c \, l_P}{2\sqrt{2}\pi} \, .
\end{eqnarray}
In the standard argument, we make the identifications
\begin{eqnarray} \label{}
(\Delta_{\psi}x)_{min} \approx \lambda_C \, ,  \ \ \ (\Delta_{\psi}p)_{max} \approx mc \, ,
\end{eqnarray} 
using $\hat{p} = -i\hbar(\partial/\partial x)$, and this gives
\begin{eqnarray} \label{}
\lambda_C \approx \frac{\hbar}{2mc} \, .
\end{eqnarray} 
However, the modified de Broglie relations (\ref{deBroglie_m>>m_P}) suggest the new identifications,
\begin{eqnarray} \label{x_max_m>>m_P}
(\Delta_{\psi}x)_{max} \approx \frac{l_P^2}{\lambda_C} = \frac{2Gm}{c^2} \, , \ \ \
(\Delta_{\psi}p')_{min} \approx mc \, .
\end{eqnarray} 
Substituting these into the modified UP (\ref{x-p_SUP_m>>m_P}) for position and momentum, and ignoring numerical factors of order unity, this becomes $m \gtrsim m_P$, in accordance with our original assumption about the validity of Eq. (\ref{x-p_SUP_m>>m_P}) in this range. Thus the identifications (\ref{x_max_m>>m_P}) are consistent with the assumption $E \gtrsim m_Pc^2$.
%
%Section3.2%%%%%%%%%%%%%%%%%%%%%%%%%%%%%%%%%%%%%%%%%%%%%%%%%%%%%%%%%%%%%%``
\subsection{de Broglie relations valid for all energies} \label{Sec.3.2}
We have assumed that the asymptotic forms of the extended de Broglie relations in the limit $E \gg m_Pc^2$ are given by Eq. (\ref{deBroglie_m>>m_P}) and showed that the mathematical structure of quantum mechanics remains unaltered by this claim, except in so far as the canonical momentum operator is multiplied by a constant factor. We now search for a unified set of expressions, valid for all energies, that reduce to Eq. (\ref{deBroglie_m>>m_P}) for $E \gg m_Pc^2$ and to the standard de Broglie relations (\ref{deBroglie}) for $E \ll m_Pc^2$. We therefore expect to recover the canonical Schr{\" o}dinger equation and expressions for $\hat{p}$ and $\hat{H}$ in the latter limit. In addition, we require all physical quantities to be continuous at $E \approx m_Pc^2$.

The simplest expressions satisfying all three conditions are 
\begin{eqnarray} \label{deBroglie_all_m}
E = \hbar\Omega, \ \ \ p = \hbar \kappa 
\end{eqnarray} 
where
\begin{equation} \label{Omega1}
\Omega = \left \lbrace
\begin{array}{rl}
\omega_P^2\left(\omega + \omega_P^2/\omega\right)^{-1} & \ (m \lesssim m_P) \\
\beta\left(\omega + \omega_P^2/\omega\right) & \ (m \gtrsim m_P) \ \ \
\end{array}\right.
\end{equation}
and
\begin{equation} \label{kappa1}
\kappa = \left \lbrace
\begin{array}{rl}
k_P^2\left(k + k_P^2/k\right)^{-1} & \ (m \lesssim m_P) \\
\beta\left(k + k_P^2/k\right) & \ (m \gtrsim m_P).
\end{array}\right.
\end{equation}
Note that continuity of $E$ and $p$, together with that of $dE/d\omega$ and $dp/dk$, at $\omega = \omega_P$ and $k=k_P$ is ensured by setting $\beta = 1/4$, but we leave it as a free constant for now. Combining Eqs. (\ref{Omega1})-(\ref{kappa1}) with Eq. (\ref{E=p^2/2m}), we must consider the different regimes of Eq. (\ref{disp_rel_all_m}) separately. 

For $m \lesssim m_P$, writing Eq. (\ref{disp_rel_all_m}) in terms of $\omega$ and $k$ gives
\begin{eqnarray} \label{disp_rel_m<m_P1*}
\frac{\hbar\omega_P^2}{(\omega + \omega_P^2/\omega)} =\frac{\hbar^2 k_P^4}{2m(k + k_P^2/k)^2} 
\end{eqnarray} 
and so
$\omega \approx (\hbar/2m)k^2$ for $\omega \ll \omega_P$ and $k \ll k_P$, as required. This may then be written as
\begin{eqnarray} \label{disp_rel_m<m_P1**}
\omega^2 - \frac{2mc^2}{\hbar k_P^2}\left(k + \frac{k_p^2}{k} \right)^2 \omega + \omega_P^2 = 0 \, ,
\end{eqnarray} 
and solved to give
\begin{eqnarray} \label{disp_rel_m<m_P1}
\omega_{\pm} = \frac{mc^2}{\hbar k_P^2}\left(k + \frac{k_P^2}{k}\right)^2\left[1 \pm \sqrt{1 - \frac{\hbar^2 k_P^6}{m^2c^2}\left(k + \frac{k_P^2}{k}\right)^{-4}} \right] \, .
\end{eqnarray} 
The reality of $\omega$ requires
\begin{eqnarray} \label{disp_rel_m<m_P2}
\left(k + \frac{k_P^2}{k}\right)^{2} \geq \frac{\hbar k_P^3}{mc} \, ,
\end{eqnarray} 
which gives a quadratic inequality in $k^2$:
\begin{eqnarray} \label{disp_rel_m<m_P3}
k^4 + \left(2k_P^2 - \frac{\hbar k_p^3}{mc} \right)k^2 + k_P^4 \geq 0 \,.
\end{eqnarray} 
This is saturated for 
\begin{eqnarray} \label{disp_rel_m<m_P4}
k_{\pm}^2 = \left(\frac{\hbar k_P^3}{2mc} - k_P^2\right)\left[1 \pm \sqrt{1 - k_P^4\left(\frac{\hbar k_P^3}{2mc} - k_P^2\right)^{-2}} \right]
\end{eqnarray} 
and the reality of $k^2$ then requires
\begin{eqnarray} \label{disp_rel_m<m_P5}
m \leq m_P' \equiv (\pi/2)m_P \, .
\end{eqnarray} 
Since the coefficient of the $k^4$ term in Eq. (\ref{disp_rel_m<m_P3}) is positive, the inequality is satisfied for $k^2 \leq k_{-}^2$ or $k^2 \geq k_{+}^2$. For $m \ll m_P'$, we may expand Eq. (\ref{disp_rel_m<m_P4}) to first order, giving
\begin{eqnarray} \label{kapprox}
k_{-} \approx \frac{mck_P}{\hbar} + \ . \ . \ . \ , \ \ \ k_{+} \approx \frac{\hbar k_P^3}{mc} + \ . \ . \ . \ .
\end{eqnarray} 
Defining $k_{\pm} = 2\pi/\lambda_{\mp}$, the lower limit on the value of wavenumber $k_{-}$ corresponds to an upper value on the wavelength $\lambda_{+}$ while the upper limit $k_{+}$ corresponds to a lower limit $\lambda_{-}$. Thus the inequality (\ref{disp_rel_m<m_P3}) is satisfied for the two wavelength ranges, 
\begin{eqnarray} \label{lambdalimit}
\lambda^2 \geq \lambda_{+}^2 \approx 2\pi(m_P/m)l_P^2 \approx \lambda_C l_P  \, , \ \ \ \lambda^2 \leq \lambda_{-}^2 \approx (2\pi)^{-1}(m/m_P)l_P^2 \approx \lambda_S l_P \, ,
\end{eqnarray} 
where the last expressions in each case neglect numerical factors of order unity. 

For $m \gtrsim m_P$, we have 
\begin{eqnarray} \label{disp_rel_m>m_P1*}
\beta\hbar(\omega + \omega_P^2/\omega) = \frac{\beta^2\hbar^2}{2m}(k + k_P^2/k)^2,
\end{eqnarray} 
which reduces to Eq. (\ref{disp_rel_m>>m_P*}) for $\omega \gg \omega_P$ and $k \gg k_P$. From here, we may perform an analysis similar to that for $m \lesssim m_P$, and the resulting formulae have the same structure as Eqs. (\ref{disp_rel_m<m_P1**})-(\ref{disp_rel_m<m_P4}), but with the substitution 
\begin{eqnarray} \label{mdual}
m \rightarrow \frac{\beta\hbar^2k_P^2}{4mc^2} = \frac{\beta\pi^2m_P^2}{m}.
\end{eqnarray} 
The reality of $k^2$ now requires
\begin{eqnarray} \label{disp_rel_m>m_P5}
m \geq 2\pi \beta m_P.
\end{eqnarray} 
Again, two discontinuous ranges of $\lambda$ are permitted:
\begin{eqnarray} \label{lambdalimit2}
\lambda^2 \geq \lambda_{+}^2 \approx (2\pi)^{-1}(4/\beta)(m/m_P)l_P^2 \approx \lambda_S l_P \, , \ \ \ \lambda^2 \leq \lambda_{-}^2 \approx 2\pi(\beta/4)(m_P/m)l_P^2 \approx \lambda_Cl_P \, . 
\end{eqnarray} 
These are equivalent to those defined by Eq.~\eqref{lambdalimit} under the interchange $\lambda_C \leftrightarrow \lambda_S$.

The continuity of $k_{\pm}^2(m)$ at $k_P$ and $\omega_{\pm}(k,m)$ at $\omega_P$ is ensured for $\beta = 1/4$ and the reality conditions on  $\omega$ and $k$ justify our initial assumptions about the division between $m \lesssim m_P$ and $m \gtrsim m_P$. For $\beta = 1/4$, Eqs.~(\ref{Omega1})-(\ref{kappa1}) become
\begin{equation} \label{Omega2}
\Omega = \left \lbrace
\begin{array}{rl}
\omega_P^2\left(\omega + \omega_P^2/\omega\right)^{-1} & \  (m \le m_P') \\ (1/4)\left(\omega + \omega_P^2/\omega\right) & \ (m \geq m_P')
\end{array}\right.
\end{equation}
and
\begin{equation} \label{kappa2}
\kappa = \left \lbrace
\begin{array}{rl}
k_P^2\left(k + k_P^2/k\right)^{-1} & \ (m \leq m_P') \\
(1/4)\left(k + k_P^2/k\right) & \ (m \geq m_P') \, ,
\end{array}\right.
\end{equation}
so that $\omega = \omega_P$, $k=k_P$ and $E = pc = \pi m_P c^2$ for $m = m_P'$. Note that our original approximate conditions, $m \lesssim m_P$ and $m \gtrsim m_P$, have now been replaced by the more precise expressions $m \leq m_P'$ and $m \geq m_P'$.

There are various dualities between the super-Planckian and sub-Planckian cases. If $\beta =1/4$, Eq.~\eqref{mdual} corresponds to the duality transformation
\begin{equation} \label{dual}
m \leftrightarrow \frac{m_P'^2}{m} \, ,
\end{equation}
which is equivalent to a reflection about the line $m = m_P'$ in the $(\log \omega, \log m)$ plane at fixed $k$. For fixed $m$, one also has the self-duality
\begin{equation} \label{self-dual_omega}
\omega_{\pm} \leftrightarrow \frac{\omega_P^2}{\omega_{\mp}} \, , 
\end{equation}
which maps the upper and lower branches of the solution $\omega_{\pm}$ into one another in the $(\log \omega, \log k)$ plane, and the self-duality 
\begin{equation} \label{self-dual_k}
k_{\pm} \leftrightarrow \frac{k_P^2}{k_{\mp}} \, . 
\end{equation}
In general, we have
\begin{equation} \label{0<m<m_P_k^2}
0 < m \leq m_P' \implies 
\left \lbrace
\begin{array}{rl}
\infty > k_{+}^2 \geq k_P^2 \\
0 < k_{-}^2 \leq k_P^2
\end{array}\right.
\end{equation}
\begin{equation} \label{m_P<m<infty_k^2}
m_P' \leq m < \infty \implies 
\left \lbrace
\begin{array}{rl}
k_P^2 \leq k_{+}^2 < \infty \\
k_P^2 \geq k_{-}^2 > 0 \, .
\end{array}\right.
\end{equation}
However, $\omega$ must be considered as a function of both $m$ and $k$, and the \emph{limiting} values of $k$ are also functions of $m$. This may be contrasted with the canonical theory, in which all modes $k \in (-\infty,\infty)$ may contribute (with some nonzero amplitude) to a given wave packet expansion.

Although $\omega \ll \omega_P$ and $k \ll k_P$, together with $m \ll m_P$ or $m \gg m_P$, give rise to the standard Schr{\" o}dinger dispersion relation or the relation proposed in Eq. (\ref{disp_rel_m>>m_P*}), respectively, these represent only one possibility in the $\omega_{-}$ branch of the full solution given by Eqs. (\ref{disp_rel_m<m_P1}) and its dual. The other possibility is $\omega \ll \omega_P$, $k \gg k_P$. We must also consider the $\omega_{+}$ solution, which is valid for both $\omega \gg \omega_P$, $k \ll k_P$ and $\omega \gg \omega_P$, $k \gg k_P$. Although it may be argued that states with $\omega > \omega_P$ or $k > k_P$ are unphysical, since they correspond to wavelengths or periods less than the Planck scales, the exact nature of the Planck scales is unknown and we cannot state definitively that they represent physical cut-offs for quantum wave modes (i.e. de Broglie waves), even if they represent genuine fundamental limits for physical observables \cite{Padmanabhan:1986ny}. Indeed, as we shall see in Sec. \ref{Sec.4}, allowing sub-Planckian wavelengths for matter waves does not lead to sub-Planckian uncertainties, $\Delta_{\psi}x < l_P$. Therefore, we here consider the formal mathematical extension of the solutions in Eq.  (\ref{disp_rel_m<m_P1}), and its dual for $m \geq m_P'$, into the sub-Planckian regimes, even if their physical interpretations are unclear. For brevity, we consider only two extremes in detail: $\omega \ll \omega_P$, $k \ll k_P$ and $\omega \gg \omega_P$, $k \gg k_P$, since the implications for the mixed regimes, $\omega \ll \omega_P$, $k \gg k_P$ and $\omega \gg \omega_P$, $k \ll k_P$, can then be easily determined. All four possible regimes are represented in Figs. 1 and 2, below.

%Fig. 1
\begin{figure}[h] \label{Fig.1}
\centering
\includegraphics[width=10cm]{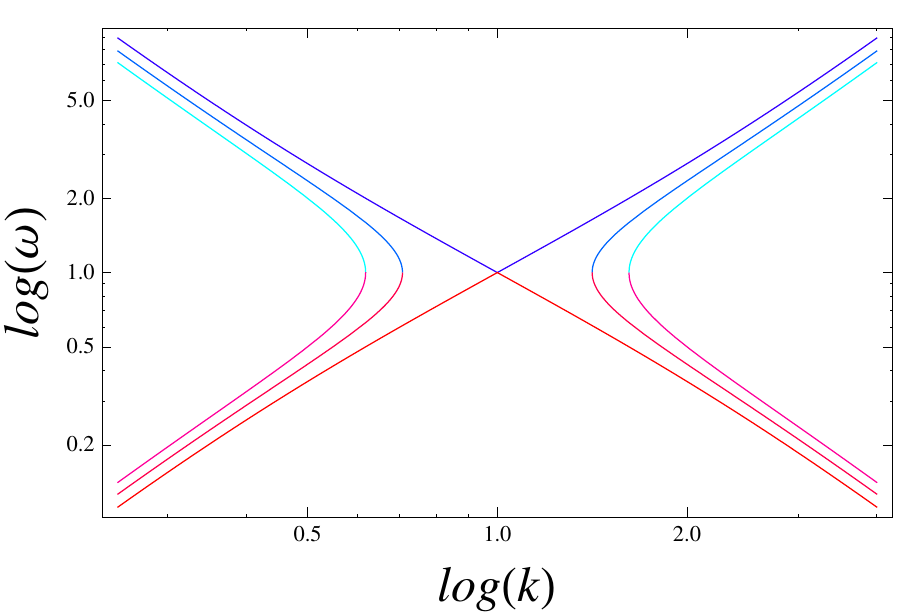}
\caption{$\omega_{+} (k:m)$ (blue) and $\omega_{-} (k:m)$ (red) as a function of $k$ for $m = m_P'$, $m_P'/(1+\pi/2)$ and $m_P'/(1+\pi)$ or, equivalently, $m_P' (1+\pi/2)$ and $m_P'(1+\pi)$
 for the non-critical cases. The standard dispersion relation of non-relativistic quantum mechanics, $\omega = (\hbar/2m)k^2$, corresponds to the asymptotic regime in the bottom left-hand corner of the diagram. In this and all subsequent plots, we choose units such that $\omega_P = k_P = 1$.}
\end{figure}

%Fig. 2
\begin{figure}[h] \label{Fig.2}
\centering
\includegraphics[width=10cm]{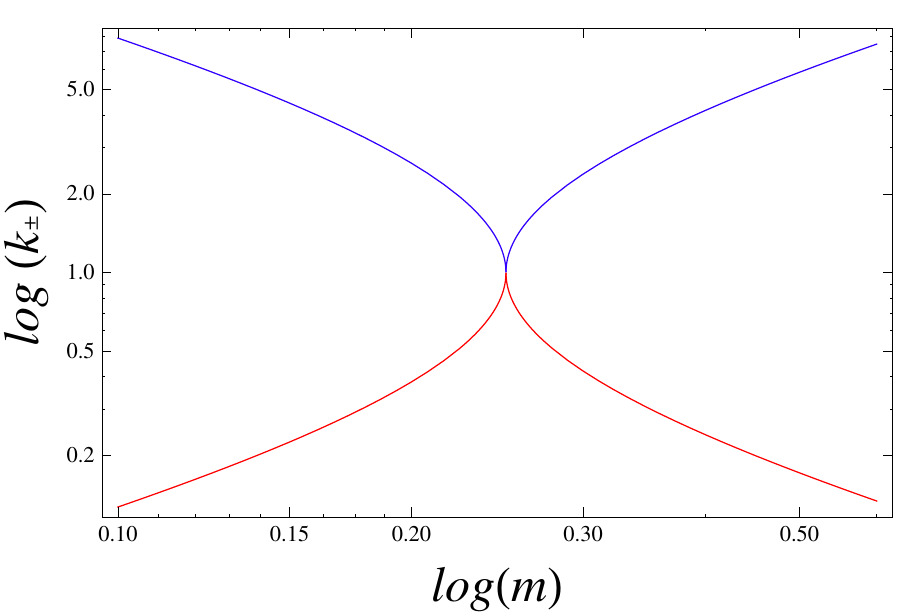}
\caption{$k_{+}$ (blue) and $k_{-}$ (red) as a function of $m$. }
\end{figure}

Let us now determine the approximate asymptotic expressions for $\omega_{\pm}$ in the various regimes. For $m \gg (\hbar k_P^3/c)(k + k_P^2/k)^{-2}$, Eq. (\ref{disp_rel_m<m_P1}) may be expanded to first order, giving
\begin{eqnarray} \label{}
\omega_{-} \approx \frac{\hbar k_P^4}{2m}\left(k + \frac{k_P^2}{k}\right)^{-2} + \ . \ . \ . \ , \ \ \ \omega_{+} \approx \frac{2mc^2}{\hbar k_P^2}\left(k + \frac{k_P^2}{k}\right)^{2} + \ . \ . \ . \ ,
\end{eqnarray}
so that 
\begin{equation} \label{m<m_P:omega-_asymp}
\omega_{-} \approx 
\left \lbrace
\begin{array}{rl}
(\hbar/2m) k^2 &  (m \gg (k^2/k_P^2)m_P, \ k \ll k_P) \\
(\hbar/2m) (k_P^4/k^2) & (m \gg (k_P^2/k^2)m_P, \ k \gg k_P)
\end{array}\right.
\end{equation}
\begin{equation} \label{m<m_P:omega+_asymp}
\omega_{+} \approx 
\left \lbrace
\begin{array}{rl}
(2mc^2/\hbar) (k_P^2/k^2) & (m \gg (k^2/k_P^2)m_P, \ k \ll k_P) \\
(2mc^2/\hbar) (k^2/k_P^2) & (m \gg \ (k_P^2/k^2)m_P, \ k \gg k_P) \, . 
\end{array}\right.
\end{equation}
For $m \ll (\beta \hbar/4k_Pc)(k + k_P^2/k)^{2}$, the duality transformation \eqref{dual} gives
\begin{eqnarray} \label{}
\omega_{-} \approx \frac{k_P^2mc^2}{8\hbar}\left(k + \frac{k_P^2}{k}\right)^{-2} + \ . \ . \ . \ , \ \ \ \omega_{+} \approx  \frac{8\hbar}{m}\left(k + \frac{k_P^2}{k}\right)^{2} + \ . \ . \ . \ ,
\end{eqnarray}
so that
\begin{equation} \label{m>m_P:omega-_asymp}
\omega_{-} \approx 
\left \lbrace
\begin{array}{rl}
(8mc^2/\hbar) (k^2/k_P^2) & (m \ll m_P(k_P^2/k^2), \ k \ll k_P)  \\ 
(8mc^2/\hbar) (k_P^2/k^2)  & (m \ll m_P(k^2/k_P^2), \ k \gg k_P) 
\end{array}\right.
\end{equation}
\begin{equation} \label{m>m_P:omega+_asymp}
\omega_{+} \approx 
\left \lbrace
\begin{array}{rl}
(\hbar/8m) (k_P^2/k^2) & (m \ll m_P(k_P^2/k^2), \ k \ll k_P) \\ 
(\hbar/8m) (k^2/k_P^2)  & (m \ll m_P(k^2/k_P^2), \ k \gg k_P) \, . 
\end{array}\right.
\end{equation}
Thus, for both $k \ll k_P$ and $k \gg k_P$, the asymptotic regions of the two solutions give $\omega_{-} \ll \omega_P$ and $\omega_{+} \gg \omega_P$ in the $m \ll m_P'$ and $m \gg m_P'$ regimes, as expected under the duality (\ref{self-dual_omega}).

The two solution branches $\omega_{\pm}(k:m)$ given by Eq. (\ref{disp_rel_m<m_P1}) are shown as functions of $k$ for $m = m_P'$, $m = m_P'/(1+\pi/2)$ and $m = m_P'/(1+\pi)$ in Fig.~1. Canonical non-relativistic quantum mechanics is recovered in the region corresponding to the bottom left, where $\omega_{-} \approx (\hbar/2m)k^2$. The branches meet at $\omega_{\pm}(k_P) = \omega_P$ for the critical case $m = m_P'$, but a gap exists in the allowed values of $k$ due to the existence of the unequal limits $k_{\pm}(m)$ for $m \neq m_P'$. By the duality (\ref{self-dual_omega}), these are equivalent to $\omega_{\pm}(k:m)$ for $m = m_P'$, $m = m_P' (1+\pi/2)$ and $m = m_P'(1+\pi)$. The limiting values of the wave-number for a given rest mass, $k_{\pm}(m)$, are plotted in Fig.~2. These obey the dualities given in Eqs. (\ref{dual}) and (\ref{self-dual_k}). 

Let us now consider the group velocity of a wave packet governed by the dispersion relations (\ref{deBroglie_all_m})-(\ref{kappa1}). In particular, we wish know its behavior at the critical point $m = m_P'$, $\omega = \omega_P$ and $k=k_P$. As this is where the upper and lower branches of $k_{\pm}$ and $\omega_{\pm}$, for both $m \leq m_P'$ and $m \geq m_P'$, unite, $\partial \omega/\partial k$ may change sign when moving between the eight difference sectors defined by $m \gtrless m_P$, $\omega \gtrless \omega_P$ and $k \gtrless k_P$. Hence, $\partial\omega/\partial k = 0$ at $m =m_P'$, $\omega = \omega_P$ and $k=k_P$ is required if continuity is to be maintained within the same branch. Differentiating Eq. (\ref{disp_rel_m<m_P1**}) gives
\begin{equation} \label{}
\frac{\partial \omega}{\partial k} = \frac{4mc^2}{\hbar k_P^3}\frac{(1-k_P^2/k^2)^2}{(1-\omega_P^2/\omega^2)}k \ \ \ (m \leq m_P')
\end{equation}
where $\omega$ is given by Eqs.~(\ref{disp_rel_m<m_P1}) and its dual under Eq. \eqref{dual} is 
\begin{equation} \label{}
\frac{\partial \omega}{\partial k} = \frac{\hbar}{4m}\frac{(1-k_P^2/k^2)^2}{(1-\omega_P^2/\omega^2)}k \ \ \ (m \geq m_P') \, .
\end{equation}
Thus if $k \rightarrow k_P$ sufficiently quickly as $\omega \rightarrow \omega_P$ and $m \rightarrow m_P'$, then $\partial\omega/\partial k = 0$ at $m = m_P'$, and is continuous. It is straightforward to verify that this is indeed the case. 

Therefore, if a given wave packet has a positive group velocity for $m < m_P'$, it will acquire a negative group velocity for $m > m_P'$, or vice-versa, depending on which solution branch we take, $\omega_{+}$ or $\omega_{-}$. We may interpret the change in sign of $\partial\omega/\partial k$ at the critical mass $m_P'$ as physically meaningful, marking a phase transition in the state of the quantum matter. In Sec. \ref{Sec.4} we will interpret transitions at $m = m_P'$, which induce a change of sign in the group velocity, as transitions between particle and black hole states. If we restrict ourselves to considering the rest frame of the wave packet centre-of-mass (which we must do in order to associate the particle momentum with the rest mass), then a change in sign of $\partial\omega/\partial k$ marks the transition between an expanding and a shrinking wave packet or, equivalently, between systems whose mass is confined within a minimum or a maximum radius. 

Finally, before moving on to consider the EOM and operators for the extended theory, we note that, acccording to Eq.~(\ref{lambdalimit}), there is a formal equivalence for $m \leq m_P'$ between $l_P$ and the step length $d$ of a random walk. In this equivalence, the limit $\lambda_{+}$ corresponds to the total distance, $\Delta = \sqrt{N}d$, where the number of steps $N$ is $ \lambda_{C}/l_P$. Random walks also give rise to Gaussian distributions in the limit $N \rightarrow \infty$, and Gaussian wave packets are the most natural models for fundamental particles/black holes. According to Eq.~(\ref{lambdalimit2}), there is an equivalence for $m \geq m_P'$ between $l_P$ and $d$ and between $\lambda_{+}$ and the total distance $\Delta$ where $N $ is $ \lambda_{S}/l_P$. The precise physical meaning of this equivalence is unclear, but it is interesting that several other approaches to unified field theories and black hole systems also employ modified commutators/dispersion relations \cite{Hinojosa:2015tga} or random walk models involving the Compton wavelength to describe quantum gravitational fluctuations \cite{Laperashvili:2015pea}. However, in many other respects, these differ greatly from the theory presented here. 
%
%Section3.3%%%%%%%%%%%%%%%%%%%%%%%%%%%%%%%%%%%%%%%%%%%%%%%%%%%%%%%%%%%%%%``
\subsection{EOM, momentum and Hamiltonian operators for the extended theory} \label{Sec.3.3}
For $m \leq m_P'$, Eq. (\ref{disp_rel_m<m_P1**}) gives
\begin{eqnarray} \label{disp_rel_m<m_P1***}
\omega^2k^2 - \frac{2mc^2}{\hbar k_P^2}\left(k^4 + 2k_P^2k^2 + k_P^4\right) \omega + \omega_P^2k^2 = 0 \, ,
\end{eqnarray} 
where the dispersion relation that gives  the standard Schr{\" o}dinger equation is recovered from the last two terms, which dominate for $\omega \ll \omega_P$ and $k \ll k_P$. Eq. (\ref{disp_rel_m<m_P1*}) suggests that the EOM for the wave function is
\begin{eqnarray} \label{Schrod_m<m_P_1}
\hat{H}\psi = \frac{\hat{p}^2}{2m}\psi = \frac{1}{2m}\frac{\hbar^2 k_P^4}{[-i(\partial /\partial x) + i k_P^2/(\partial /\partial x)]^2}\psi = \frac{\hbar \omega_P^2}{[i(\partial /\partial t) - i \omega_P^2/(\partial /\partial t)]}\psi \, .
\end{eqnarray} 
We may rewrite this expression as 
\begin{eqnarray} \label{H:m<m_P:w<w_p:k<k_P}
\hat{H}\psi = \frac{\hat{p}^2}{2m}\psi = -\frac{\hbar^2}{2m}[1 - k_P^{-2}(\partial^2 /\partial x^2)]^{-2}\frac{\partial^2 \psi}{\partial x^2} = i\hbar[1 - \omega_P^{-2}(\partial^2 /\partial t^2)]^{-1}\frac{\partial \psi}{\partial t}
\end{eqnarray} 
and define the operators in square brackets as Taylor series, 
\begin{eqnarray} \label{Exp1A}
[1 - k_P^{-2}(\partial^2 /\partial x^2)]^{-2} = \sum_{n=0}^{\infty}(n+1)k_P^{-2n}\frac{\partial^{2n}}{\partial x^{2n}} \, , 
\end{eqnarray} 
\begin{eqnarray} \label{Exp1B}
[1 - \omega_P^{-2}(\partial^2 /\partial t^2)]^{-1} = \sum_{n=0}^{\infty}\omega_P^{-2n}\frac{\partial^{2n}}{\partial t^{2n}} \, .
\end{eqnarray} 
Technically, these expansions are valid when applied to wave packets composed of plane-wave modes, $\phi(k,\omega:x,t) = \exp[i(kx-\omega t)]$ with $\omega < \omega_P$ and $k < k_P$, since
\begin{eqnarray} \label{Exp1C}
\sum_{n=0}^{\infty}(n+1)k_P^{-2n}\frac{\partial^{2n}\phi}{\partial x^{2n}} &=& \sum_{n=0}^{\infty}(n+1)k_P^{-2n}(ik)^{2n}\phi = [1 + (k/k_P)^2]^{-2}\phi \, , 
\end{eqnarray} 
\begin{eqnarray} \label{Exp1D}
\sum_{n=0}^{\infty}\omega_P^{-2n}\frac{\partial^{2n}\phi}{\partial t^{2n}}  &=& \sum_{n=0}^{\infty}\omega_P^{-2n}(-i\omega)^{2n}\phi = [1 + (\omega/\omega_P)^2]^{-1}\phi \, ,
\end{eqnarray} 
and the radius of convergence for the series are given by $k/k_P < 1$ and $\omega/\omega_P < 1$. In the $\omega \ll \omega_P$ and $k \ll k_P$ regime, the higher order terms are subdominant and we obtain the standard Schr{\" o}dinger equation. However, for $\omega \approx \omega_P$ and $k \approx k_P$, which applies \emph{necessarily} for any particle with $m \approx m_P'$, the higher order corrections become significant. 

This is an important point. Even in the standard theory, we are not free to consider a quantum state with $\omega \approx \omega_P$ and $k \approx k_P$ for \emph{any} $m$, since these values necessarily imply $E \approx 2\pi m_Pc^2$, $m \approx \pi m_P$, by substitution of the usual de Broglie relations into the non-relativisitc energy-momentum relation. Similarly, in the extended theory, the limit $m \rightarrow m_P'$ implies $\omega \rightarrow \omega_P$, $k \rightarrow k_P$ and vice-versa. This point is explicitly inaccessible for particles with $m \neq m_P'$, due to the existence of the limiting values $k_{\pm}$. Thus, systems with mass $m_P'$, be they interpreted as particles or black holes, are \emph{unique} in this prescription. Alternatively, such systems may be interpreted as black hole relics formed at the end of Hawking evaporation. These relics represent unique quantum states, that fulfill the requirements of both fundamental particles and black holes, where the Schwarzschild and Compton radii coincide (cf. \cite{cmn}).

Since the series (\ref{Exp1C})-(\ref{Exp1D}) do not converge for $\omega \geq \omega_P$ and $k \geq k_P$, this motivates an alternative definition of the Hamiltonian and momentum operators in the sub-Planckian regime, based on rearranging Eq. (\ref{Schrod_m<m_P_1}) as
\begin{eqnarray} \label{H:m<m_P:w>w_p:k>k_P}
\hat{H}'\psi = \frac{\hat{p}'^2}{2m}\psi = -4mc^2k_P^{-2}[1- \omega_P^2/(\partial^2/\partial t^2)]^{-1}\frac{\partial^2 \psi}{\partial x^2} = 2i\hbar[1- k_P^2/(\partial^2/\partial x^2)]^{-2}\frac{\partial \psi}{\partial t} \, .
\end{eqnarray}
Here primes indicate modified Hamiltonian and momentum operators, which apply on sub-Planckian scales (i.e. that operate on plane-wave modes with $\omega \geq \omega_P$, $k \geq k_P$). The terms in square brackets are defined via
\begin{eqnarray} \label{Exp2A}
[1 - k_P^{2}/(\partial^2 /\partial x^2)]^{-2} = \sum_{n=0}^{\infty}(n+1)k_P^{-(2n+4)}\frac{\partial^{(2n+4)}}{\partial x^{(2n+4)}} \, , 
\end{eqnarray} 
\begin{eqnarray} \label{Exp2B}
[1 - \omega_P^{2}/(\partial^2 /\partial t^2)]^{-1} = -\sum_{n=0}^{\infty}\omega_P^{-(2n+2)}\frac{\partial^{(2n+2)}}{\partial t^{(2n+2)}} \, .
\end{eqnarray} 
In this case, we have 
\begin{eqnarray} \label{Exp2C}
\sum_{n=0}^{\infty}(n+1)k_P^{-(2n+4)}\frac{\partial^{(2n+4)}\phi}{\partial x^{(2n+4)}} &=& \sum_{n=0}^{\infty}(n+1)k_P^{-(2n+4)}(ik)^{(2n+4)}\phi = [1 + (k_P/k)^2]^{-2}\phi, 
\end{eqnarray} 
\begin{eqnarray} \label{Exp2D}
-\sum_{n=0}^{\infty}\omega_P^{-(2n+2)}\frac{\partial^{(2n+2)}\phi}{\partial t^{(2n+2)}} &=& -\sum_{n=0}^{\infty}\omega_P^{-(2n+2)}(-i\omega)^{(2n+2)}\phi = [1 + (\omega_P/\omega)^2]^{-1}\phi \, . 
\end{eqnarray} 
Strictly speaking, the series on left-hand sides of Eqs. (\ref{Exp2C}) and (\ref{Exp2D}) only converge to the expressions on the right-hand sides for $k/k_P <1$ and $\omega/\omega_P <1$, respectively, as before. However, the expressions on the right may also be expanded as
\begin{eqnarray} \label{Exp2C*}
[1 + (k_P/k)^2]^{-2} = \sum_{n=0}^{\infty}(n+1)k^{-2n}(ik_P)^{2n},
\end{eqnarray} 
\begin{eqnarray} \label{Exp2D*}
[1 + (\omega_P/\omega)^2]^{-1} = \sum_{n=0}^{\infty}\omega^{-2n}(-i\omega_P)^{2n},
\end{eqnarray} 
which are analogous to the expressions in Eqs. (\ref{Exp1C})-(\ref{Exp1D}), but with $k \leftrightarrow k_P$ and $\omega \leftrightarrow \omega_P$ interchanged. This motivates the definitions
\begin{eqnarray} \label{Exp2A*}
[1 - k_P^{2}/(\partial^2 /\partial x^2)]^{-2} = \sum_{n=0}^{\infty}(n+1)k_P^{2n}\left(\frac{\partial^{2n}}{\partial x^{2n}}\right)^{-1} \, , 
\end{eqnarray} 
\begin{eqnarray} \label{Exp2B*}
[1 - \omega_P^{2}/(\partial^2 /\partial t^2)]^{-1} = \sum_{n=0}^{\infty}\omega_P^{2n}\left(\frac{\partial^{2n}}{\partial t^{2n}}\right)^{-1} \, ,
\end{eqnarray} 
where it is understood that 
\begin{eqnarray} \label{understand}
\left(\frac{\partial^m}{\partial x^m}\right)^{-1}\phi = (ik)^{-m}\phi, \ \ \ \left(\frac{\partial^m}{\partial t^m}\right)^{-1}\phi = (-i\omega)^{-m}\phi, 
\end{eqnarray} 
for all $m \in \mathbb{N}$. Thus, acting on momentum eigenstates $\phi$, the expansions (\ref{Exp2A*})-(\ref{Exp2B*}) converge for $\omega > \omega_P$, $k>k_P$. 
\\
\indent
In the limit $\omega \gg \omega_P$, $k \gg k_P$, Eq. (\ref{H:m<m_P:w>w_p:k>k_P}) reduces to
\begin{eqnarray} \label{Schrod_m<<m_P_2*}
\hat{H}'\psi = \frac{\hat{p}'^2}{2m}\psi \approx -\frac{4mc^2}{k_P^2}\frac{\partial^2 \psi}{\partial x^2} \approx 2i\hbar\frac{\partial \psi}{\partial t} \, ,
\end{eqnarray}
This is equivalent to the EOM (\ref{Schrod_1D}), obtained in the limit $\omega \ll \omega_P$ and $k \ll k_P$ for $m \gg m_P,$ up to numerical factors of order unity. With this choice of factors, however, Eqs. (\ref{H:m<m_P:w<w_p:k<k_P}) and  (\ref{H:m<m_P:w>w_p:k>k_P}) give $\hat{H}\psi = \hat{H}'\psi = 2 m_P'c^2$ for $m = m_P'$, $\omega = \omega_P$, $k = k_P$, via an appropriate limiting procedure. 
\\
\indent
For $m \geq m_P'$, the same analysis applies providing one replaces $m$ with $m_P'^2/m$ in all the above equations. However, we here treat it in detail, since some of the ideas, such as redefining Hamiltonians (motivated by extended dispersion relations), are unfamiliar. The dual of Eq. (\ref{disp_rel_m<m_P1**}) under Eq. (\ref{dual}) gives
\begin{eqnarray} \label{disp_rel_m>m_P1***}
\omega^2k^2 - \frac{\hbar}{8m}\left(k^4 + 2k_P^2k^2 + k_P^4\right) \omega + \omega_P^2k^2 = 0,
\end{eqnarray} 
and Eq. (\ref{disp_rel_m>m_P1*}) suggests the EOM 
\begin{eqnarray} \label{Schrod_m>m_P_1}
\hat{H}\psi = \frac{\hat{p}^2}{2m}\psi = \frac{\hbar^2}{32m}[-i(\partial /\partial x) + i k_P^2/(\partial /\partial x)]^2\psi = \frac{\hbar}{4} [i(\partial /\partial t) - i \omega_P^2/(\partial /\partial t)]\psi \, .
\end{eqnarray} 
We can make sense of this expression for $\omega < \omega_P$, $k < k_P$ by rewriting it as
\begin{eqnarray}  \label{H:m>m_P:w<w_p:k<k_P}
\hat{H}'\psi = \frac{\hat{p}'^2}{2m}\psi = -mc^2k_P^{-2}[1 -  \omega_P^{-2}(\partial^2 /\partial t^2)]\frac{\partial^2 \psi}{\partial x^2} = \frac{i\hbar}{8} [1 - k_P^{-2} (\partial^2 /\partial x^2)]^2\frac{\partial \psi}{\partial t} \, ,
\end{eqnarray} 
which, in the $\omega \ll \omega_P$ and $k \ll k_P$ limit, reduces to 
\begin{eqnarray}  \label{}
\hat{H}'\psi = \frac{\hat{p}'^2}{2m}\psi \approx -mc^2k_P^{-2}\frac{\partial^2 \psi}{\partial x^2} = \frac{i\hbar}{8}\frac{\partial \psi}{\partial t} \, ,
\end{eqnarray} 
which is analogous to Eq. (\ref{Schrod_m<<m_P_2*}) up to numerical factors. Our previous results then motivate the definition
\begin{eqnarray}\label{H:m>m_P:w>w_p:k>k_P} 
\hat{H}\psi = \frac{\hat{p}^2}{2m}\psi =-\frac{\hbar^2}{32m} \left[1 - \frac {k_P^{2}}{\partial^2 /\partial x^2}\right]^2\frac{\partial^2 \psi}{\partial x^2} = \frac{i\hbar}{4} \left[1 - \frac{\omega_P^{2}}{ \partial^2 /\partial t^2} \right]\frac{\partial \psi}{\partial t} \, ,
\end{eqnarray} 
for $\omega > \omega_P$ and $k > k_P$, where we can make sense of the terms in square brackets using Eq. (\ref{understand}). Eq. (\ref{H:m>m_P:w>w_p:k>k_P}) reduces to
\begin{eqnarray}\label{}
\hat{H}\psi = \frac{\hat{p}^2}{2m}\psi \approx -\frac{\hbar^2}{32m}\frac{\partial^2 \psi}{\partial x^2} \approx \frac{i\hbar}{4}\frac{\partial \psi}{\partial t}
\end{eqnarray} 
in the limit  $\omega \gg \omega_P$ and $k \gg k_P$, i.e. to the standard Schr{\" o}dinger equation up to numerical factors. It is again straightforward to check that Eqs.~(\ref{H:m>m_P:w<w_p:k<k_P}) and  (\ref{H:m>m_P:w>w_p:k>k_P}) give $\hat{H}\psi = \hat{H}'\psi = 2m_P'c^2$ for $m = m_P'$, $\omega = \omega_P$ and $k = k_P$, by taking appropriate limits. 
\\
\indent
By comparison with Eqs. (\ref{m<m_P:omega-_asymp})-(\ref{m<m_P:omega+_asymp}) and (\ref{m>m_P:omega-_asymp})-(\ref{m>m_P:omega+_asymp}), we see that redefining the momentum and Hamiltonian operators for $m \leq m_P'$ corresponds to taking the $\omega_{+}$ branch of the solution, whereas the original definitions correspond the $\omega_{-}$ branch. For $m \geq m_P'$, the reverse is true, and the redefinition corresponds to taking the $\omega_{-}$ branch, whereas the original definitions correspond to $\omega_{+}$. Hence the EOM in the asymptotic regions $m \leq m_P'$, $\omega \ll \omega_P$, $k \ll k_P$ and $m \geq m_P'$,  $\omega \gg \omega_P$, $k \gg k_P$ are nearly the same, differing only up to numerical factors. The same applies for  the asymptotic regions $m \geq m_P'$, $\omega \ll \omega_P$, $k \ll k_P$ and $m \leq m_P'$,  $\omega \gg \omega_P$, $k \gg k_P$. This symmetry is indicated in Fig. 1 and results from the fact that the asymptotes in opposite corners of the $(\log \ \omega, \log \ k)$ diagram are parallel.
\\
\indent
As we shall see in the next section, a similar correspondence holds between the critical wavelengths associated with the maximum or minimum positional uncertainty of the system, $\lambda_{crit}^{\pm}(m)$. These determine the modified Compton and Schwarzschild lines in the extended de Broglie theory. The $\lambda_{crit}^{+}$ lines for $m \leq m_P'$, $\omega \ll \omega_P$, $k \ll k_P$ and $m \geq m_P'$, $\omega \ll \omega_P$, $k \ll k_P$ are equivalent to the $\lambda_{crit}^{-}$ lines for $m \geq m_P'$,  $\omega \gg \omega_P$, $k \gg k_P$ and $m \leq m_P'$,  $\omega \gg \omega_P$, $k \gg k_P$, respectively. Between the two half-planes, $m \leq m_P'$ and $m \geq m_P'$, the modified Compton and Schwarzschild lines (denoted by $\lambda_C'$ in the first region and $\lambda_S'$ in the second) form a continuous curve, with the separation between particle and black hole states appearing at $m = m_P'$, where the $\lambda_{crit}^{\pm}$ curves from all four sectors meet. Crucially, $\lambda_C'$ may be interpreted as a minimum radius, whereas $\lambda_S'$ is interpreted as a maximum radius. Both curves remain \emph{above} the line $\lambda = l_P$, suggesting that physically observable length scales remain super-Planckian, even if we allow for the existence of sub-Planckian de Broglie waves in the extended theory.
%
%%%%%%%%%%%%%%%%%%%%%%%%%%%%%%%%%%%%%%%%%%%%%%%%%%%%%%%%%%%%%%%%%%%
%Section4%%%%%%%%%%%%%%%%%%%%%%%%%%%%%%%%%%%%%%%%%%%%%%%%%%%%%%%%%%%%%%``   
\section{Compton$-$Schwarzschild correspondence from extended de Broglie relations} \label{Sec.4}
We have demonstrated the consistency of the redefined momentum operator (\ref{p_x_1D}) in the asymptotic limit $m \gg m_P$, $\omega \ll \omega_P$, $k \ll k_P$ with the uncertainty relations via the identifications in Eq. (\ref{x_max_m>>m_P}). We may do the same with the redefined momentum operator (\ref{H:m<m_P:w>w_p:k>k_P}) in the limit $m \ll m_P$, $\omega \gg \omega_P$, $k \gg k_P$ by again identifying $\Delta_{\psi}x$ and $\Delta_{\psi}p'$ with appropriate quantities. However, it is easier to work directly with Eqs.~(\ref{Schrod_m<m_P_1}) and (\ref{Schrod_m>m_P_1}) in order to obtain a phenomenological model which is consistent with our previous results. Equation (\ref{Schrod_m<m_P_1}) suggests the identifications
\begin{eqnarray}\label{Ident1}
\Delta_{\psi}x \approx k_1\left(c_1'\lambda_{crit} + \frac{l_P^2}{c_1' \lambda_{crit}}\right) = \left(c_1\lambda_{crit} + \frac{c_2l_P^2}{ \lambda_{crit}}\right) \, , \ \ \ 
\Delta_{\psi}p \approx  m_Pc \mu^{-1}  \ \ \ (m \leq m_P')  
\end{eqnarray} 
where 
\begin{eqnarray}
 \mu \equiv \frac{m}{m_P'} + \frac{m_P'}{m} \, .
\label{mu}
\end{eqnarray}
Eq.~(\ref{Schrod_m>m_P_1}) then gives
\begin{eqnarray}\label{Ident2}
\Delta_{\psi}x \approx k_2 l_P^2\left(c_2'\lambda_{crit} + \frac{l_P^2}{c_2'\lambda_{crit}}\right)^{-1} = l_P^2\left(c_3\lambda_{crit} + \frac{c_4l_P^2}{\lambda_{crit}}\right)^{-1} \, , \ \ 
\Delta_{\psi}p \approx  m_P c \frac{\mu}{4} \ \ (m \geq m_P') \, ,
\end{eqnarray} 
where in both cases
\begin{eqnarray}\label{}
\hat{p} \approx -i\hbar(\partial/\partial x) \, , \ \ \ [\hat{x},\hat{p}] \approx i\hbar \, ,
\end{eqnarray} 
to first order. The  numerical factors $c_1-c_4$ must be fixed by requiring these formulae to reduce to the correct asymptotic forms and by matching $\lambda_{crit}^{\pm}$ at $m=m_P'$. Note that there is always some ambiguity, even in the standard theory, as to the constants of proportionality between $\Delta_{\psi}x$ and $\lambda_C$ or $\Delta_{\psi}p$ and $mc$. In the extended theory, what is important is that the free constants are fixed so as to give the correct asymptotic expressions for $\lambda_C$ or $\lambda_S$.
\\
\indent
The continuity of $\Delta_{\psi}p$ at the critical point is ensured by the definitions above, since we have chosen to absorb all the ambiguity associated with the uncertainties into the identification of $\Delta_{\psi}x$. Thus, if a self-consistent formulation is possible, fixing the values of the constants $c_1-c_4$ and imposing continuity of $\lambda_{crit}$ at $m=m_P'$ should also ensure continuity of $\Delta_{\psi}x$ at some length scale $\Delta_{\psi}x = \gamma l_P$ with $\gamma > 0$. This implies 
\begin{eqnarray}\label{alpha}
c_1 + c_2 = \frac{\gamma}{c_3 + c_4} \, .
\end{eqnarray} 
We note also that the numerical factor multiplying the standard expression for the Compton wavelength (interpreted as the point at which pair-production become significant) is to some degree arbitrary. We now fix this by requiring the Compton line to intersect the Schwarzschild line at $m=m_P'$. Putting
\begin{eqnarray}\label{}
\lambda_C = \zeta \frac{l_pm_P}{2m} = \lambda_S = \frac{2l_Pm}{m_P} 
\end{eqnarray} 
implies
\begin{eqnarray}\label{}
\zeta = 4m^2/m_P^2 \, .
\end{eqnarray} 
Then substituting $m = m_P'$ gives
\begin{eqnarray}\label{}
\zeta = \pi^2 \, 
\end{eqnarray}
and this implies the correspondence
\begin{eqnarray}\label{X}
\lambda_C =  \frac{\pi^2 l_pm_P}{2m} \leftrightarrow \lambda_S = \frac{2l_Pm}{m_P} 
\end{eqnarray} 
under the transformation $m \leftrightarrow m_P'^2/m$. 
\\
\indent
From the UP and Eq.~(\ref{Ident1}), we have 
\begin{eqnarray}\label{}
c_1\lambda_{crit}^2 - \frac{1}{2} \mu l_P \lambda_{crit} + c_2l_P^2 \geq 0
\end{eqnarray}
for $m \leq m_P'$, which gives the solution
\begin{eqnarray}\label{lambda_crit^pm:m<m_P}
\lambda_{crit}^{\pm} = \frac{l_P \mu}{4c_1} \left[1 \pm \sqrt{1 - 16c_1c_2 \mu^{-2}} \right],
\end{eqnarray}
where the inequality is satisfied for $\lambda \geq \lambda_{crit}^{+}$ or $\lambda \leq \lambda_{crit}^{-}$. In the limit $m \ll m_P'$, this gives
\begin{eqnarray}\label{lambda_crit_m<m_P}
\lambda_{crit}^{+} \approx (\pi/4)c_1^{-1}(m_P/m)l_P +  \ . \ . \ . \ , \ \ \ \lambda_{crit}^{-} \approx (4/\pi)c_2(m/m_P)l_P +  \ . \ . \ . \  .
\end{eqnarray}
Since we require 
$\lambda_{crit}^{+} = \lambda_C = (\pi m_P'/m)l_P$ 
in this limit, this sets
\begin{eqnarray}\label{}
c_1 = (2\pi)^{-1}.
\end{eqnarray} 
\indent
From the UP and Eq. (\ref{Ident2}), we have 
\begin{eqnarray}\label{}
c_3\lambda_{crit}^2 - \frac{1}{2} \mu l_P \lambda_{crit} + c_4l_P^2 \leq 0
\end{eqnarray}
for $m \geq m_P'$. This gives the solution
\begin{eqnarray}\label{lambda_crit^pm:m>m_P}
\lambda_{crit}^{\pm} = \frac{l_P \mu }{4c_3} \left[1 \pm \sqrt{1 - 16c_3c_4 \mu^{-2}}  \right],
\end{eqnarray}
where the inequality is satisfied for $\lambda_{crit}^{-} \leq \lambda \leq \lambda_{crit}^{+}$. In the limit $m \gg m_P'$, this gives
\begin{eqnarray}\label{lambda_crit_m>m_P}
\lambda_{crit}^{+} \approx \pi^{-1}c_3^{-1}(m/m_P)l_P +  \ . \ . \ . \ , \ \ \ \lambda_{crit}^{-} \approx \pi c_4(m_P/m)l_P +  \ . \ . \ . \  .
\end{eqnarray}
Since we require $\lambda_{crit}^{+} = \lambda_S = 2(m/m_P)l_P$ in this limit, this sets
\begin{eqnarray}\label{}
c_3 = c_1 = (2\pi)^{-1}.
\end{eqnarray} 
The continuity of $\lambda_{crit}^{\pm}$ at $m = m_P'$ requires $c_1c_2 = c_3c_4$, and hence $c_2 = c_4$. If we impose the further condition $\lambda_{crit}^{+} = \lambda_{crit}^{-}$ at the critical point, by analogy with our previous results, this implies
\begin{eqnarray}\label{}
c_2 = c_4 = (\pi/2).
\end{eqnarray} 
Substituting for $c_1$ to $c_4$ in Eq. (\ref{alpha}) then implies $\gamma =1$, so that $\Delta_{\psi}x = l_P$ at $m = m_P'$ and is continuous, as required. 
Thus we have 
\begin{equation} \label{}
\Delta_{\psi}x 
\left \lbrace
\begin{array}{rl}
\gtrsim (\Delta_{\psi}x)_{min} =& [\lambda_{crit}^{\pm}/(2\pi) + (\pi/2)l_P^2/\lambda_{crit}^{\pm}]  \ \ \  (m \leq m_P') \\
\lesssim (\Delta_{\psi}x)_{max} =& [\lambda_{crit}^{\pm}/(2\pi) + (\pi/2)l_P^2/\lambda_{crit}^{\pm}] \ \ \ (m \geq m_P') \, ,
\end{array}\right.
\end{equation}
where the $\lambda_{crit}^{\pm}$ are given by Eq. (\ref{lambda_crit^pm:m<m_P}) or (\ref{lambda_crit^pm:m>m_P}), since these define the same curve after fixing the values of $c_1-c_4$. 
The value of $\Delta_{\psi}x$ is independent of the solution branch, since 
\begin{eqnarray}\label{}
\lambda_{crit}^{\pm}/(2\pi) \leftrightarrow (\pi/2)l_P^2/\lambda_{crit}^{\mp} \, .
\end{eqnarray} 
Although there are only two curves, $\lambda_{crit}^{\pm}$, which are valid for all values of $m$, the direction of the inequality implied by the UP changes at $m = m_P'$, so that the allowed wavelength ranges for a given mass also change. Specifically, for $m \leq m_P'$, the UP inequality is satisfied for $\lambda_{crit} \geq \lambda_{crit}^{+}$ or $\lambda_{crit} \leq \lambda_{crit}^{-}$ whereas, for  $m \geq m_P'$,  it is satisfied for $\lambda_{crit}^{-} \leq \lambda_{crit} \leq \lambda_{crit}^{+}$. 
\\
\indent
A plot of the standard Compton line, defined as $\lambda_C = (\pi m_P'/m)l_P$, and Schwarzschild line is given in Fig.~3. Both curves, $\lambda_{crit}^{\pm}$, are shown in Fig.~4. The lines in the former are straight, whereas those in the latter are $V$-shaped. In Fig.~3, the Compton line may be trusted only up to $m = m_P'$ since gravitational effects become important above this. Likewise, the standard Schwarzschild line may be trusted only down to this point since  quantum effects become significant below this. This effectively rules out the lower half of the diagram, which corresponds to de Broglie waves with sub-Planckian wavelengths and  periods. Thus, in the standard picture, we must rely on non-gravitational quantum mechanics for $m \leq m_P'$ and classical gravity for $m \geq m_P'$, and the lower half plane is completely inaccessible in either case. By contrast, in Fig. 4, $\lambda_{crit}^{+}$ covers both the gravitational and non-gravitational sectors (i.e. the regions corresponding to both fundamental particles and black holes) in the super-Planckian regime, whereas $\lambda_{crit}^{-}$ covers both sectors in the sub-Planckian regime.
%
%Fig. 3
\begin{figure}[h] \label{Fig.3}
\centering
\includegraphics[width=12cm]{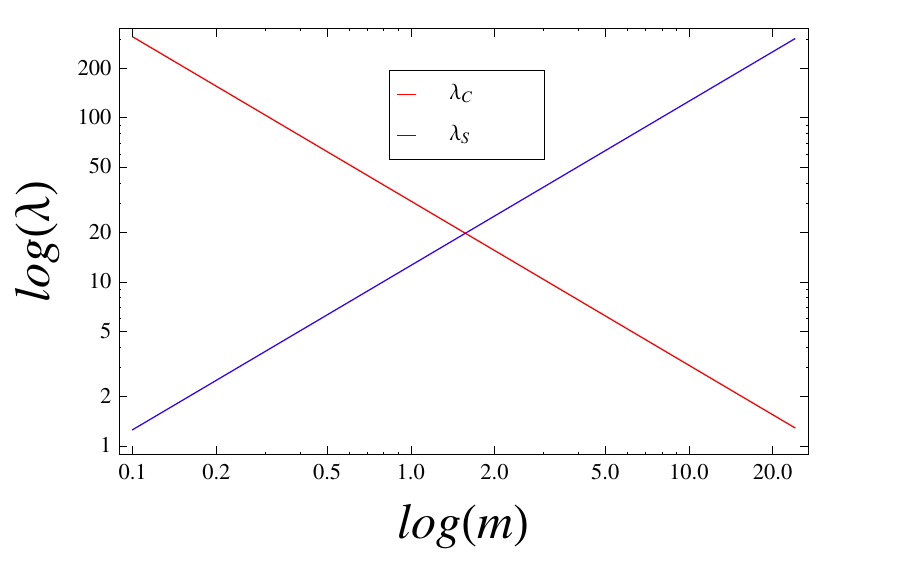}
\caption{Showing the Compton line in canonical non-relativistic quantum mechanics, $\lambda_C = (\pi m_P'/m)l_P$, the Schwarzschild line, $\lambda_{S} = 2(m_P/m)l_P$, and their intersect at $m =  m_P'$.}
\end{figure}
%
%Fig. 4
\begin{figure}[h] \label{Fig.4}
\centering
\includegraphics[width=12cm]{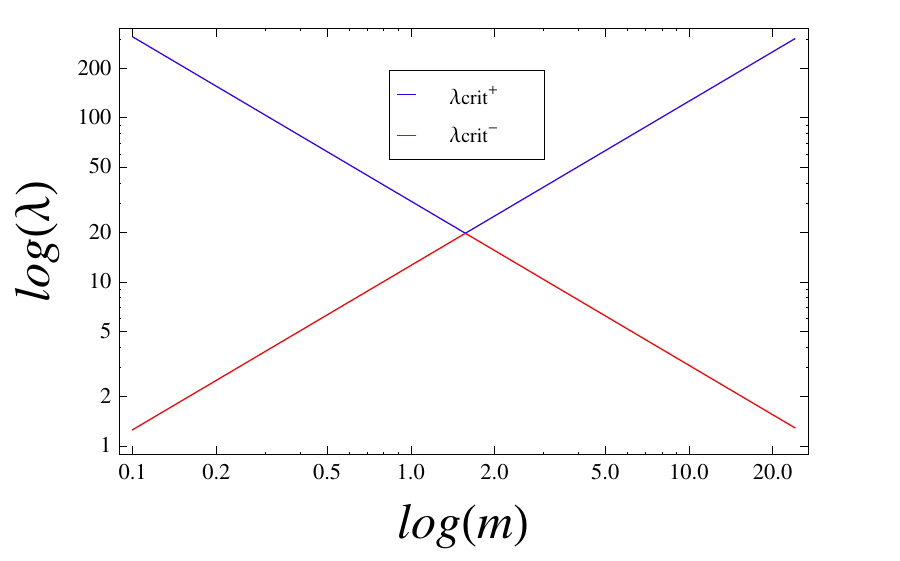}
\caption{$\lambda_{crit}^{\pm}$, obtained from Eqs. (\ref{lambda_crit^pm:m<m_P})/(\ref{lambda_crit^pm:m>m_P}) for $c_3 = c_1 = (2\pi)^{-1}$ and $c_2 = c_4 = \pi/2$.}
\end{figure}

In order to recover the usual identifications 
\begin{eqnarray}
(\Delta_{\psi}x)_{min} \approx \lambda_C = (1/2)l_P(m_P/m) \, , \ \ \ (\Delta_{\psi}p)_{max} \approx mc
\end{eqnarray}
in the limit $m \ll  m_P'$, it is clear that we require
\begin{eqnarray}\label{4.22}
\lambda'_{C/S} = 2\pi \Delta_{\psi}x = 2\pi [(\lambda_{crit}^{\pm}/(2\pi)) + ((\pi/2)l_P^2/(\lambda_{crit}^{\pm})], \ \ \ mc = (\pi/2) \Delta_{\psi}p \, .
\end{eqnarray} 
Here $\lambda'_{C/S}$ denotes a single curve but we use the notation $\lambda'_{C} = 2\pi(\Delta_{\psi}x)_{min}$ for $ m \leq m_P'$ and $\lambda'_{S} = 2\pi(\Delta_{\psi}x)_{max}$ for $m \geq m_P'$, in order to distinguish the physical implications of each limit: one represents a minimum spatial width for the wave packet, whereas the other represents a maximum. There is no difficulty in identifying $2\pi\Delta_{\psi}x$ with a given limit rather than $\Delta_{\psi}x$, since the uncertainty represents an average over all plane wave modes in the wave packet $\psi$, not an absolute maximum/minimum attainable value. Thus, we have
\begin{eqnarray}\label{}
\lambda'_{C} \approx (\pi m_P'/m)l_P \ \ \ (m \ll m_P'), \ \ \
\lambda'_{S} \approx 2(m_P/m)l_P \ \ \  (m \gg m_P') \, 
\end{eqnarray} 
in the asymptotic regions, but the two lines meet smoothly at $m = m_P'$, $\lambda'_{C} = \lambda'_{S} = 2\pi l_P$. Fig.~5 shows the $\lambda'_{C}$ and $\lambda'_{S}$ curve, together with the $\lambda_{C}$ and $\lambda_{S}$ lines to which it tends for $m \ll m_P'$ and $m \gg m_P'$. Fig. 6 shows $\lambda'_{C}$ and $\lambda'_{S}$, together with the shaded regions that represent the direction of the inequality obtained from the UP.
%
%Fig. 5
\begin{figure}[h] \label{Fig.5}
\centering
\includegraphics[width=12cm]{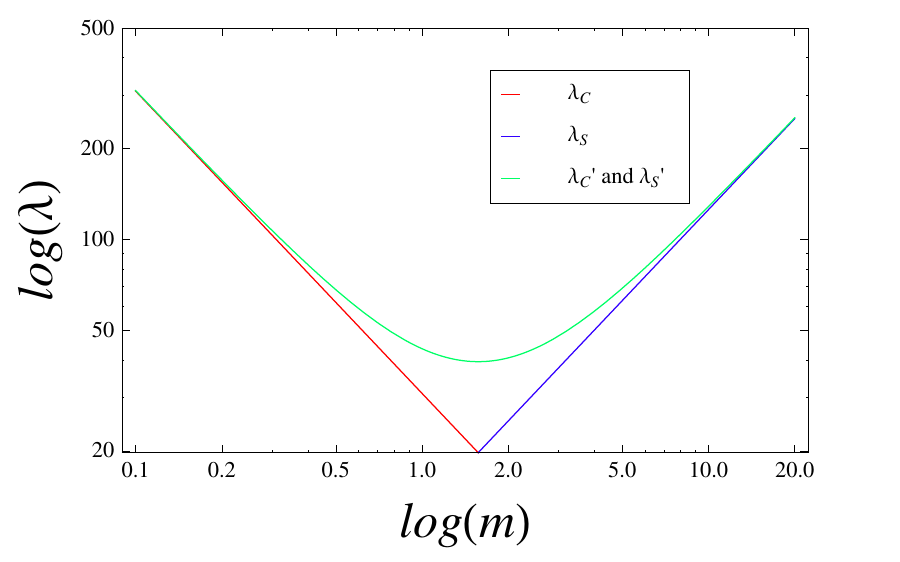}
\caption{Comparing $\lambda'_{C/S}$ from Eq. (\ref{4.22}) (green curve) to the asymptotes $\lambda_{C} = (\pi m_P'/m)l_P$ (red line) and $\lambda_{S} = 2(m_P/m)l_P$ (blue line).}
\end{figure}
%
%Fig. 6
\begin{figure}[h] \label{Fig.6}
\centering
\includegraphics[width=11cm]{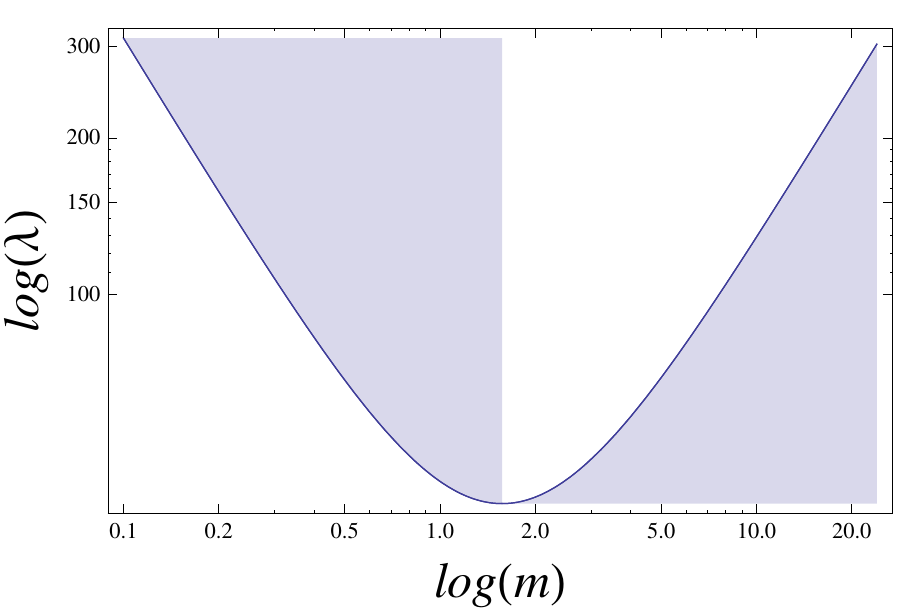}
\caption{The $\lambda'_{C/S}$ curve, including the shaded regions representing the allowed regions according to the direction of the inequality obtained from the UP. We refer to the generalized Compton line, $\lambda'_{C}$, for $ m \leq m_P'$ and the generalized event horizon, $\lambda'_{S}$, for $ m \geq m_P'$.}
\end{figure}
\\ \indent 
In Schr{\" o}dinger's equation, $(\hbar/2m)$ is formally equivalent to a diffusion coefficient for the complex `density' $\psi$. By rewriting the equation as $i(\partial \psi/\partial ct) = (\hbar/2mc)\vec{\nabla}^2\psi$, we may interpret the standard expression for the Compton wavelength as giving the order of magnitude value of the corresponding diffusion length. Physically, for a particle with rest mass $m \geq m_P'$ (i.e. a black hole), the two solution branches $\lambda_{\mp} = (2\pi)/k_{\pm}$ obtained from the dual of Eq. (\ref{disp_rel_m<m_P4}), may be interpreted as limits on the diffusion lengths for a wave packet with self-gravity. By contrast, the two solutions $\lambda_{crit}^{\pm}$ obtained from Eq.~(\ref{lambda_crit^pm:m>m_P}) --  or, more strictly, the combination $\lambda_{C/S}' $ -- represent maximum and minimum bounds on the spatial localization of the wave packet, obtained by performing measurements to ascertain the position of the centre of mass. In canonical quantum mechanics, the characteristic diffusion and localization lengths coincide, whereas in the modified theory presented here they are related, by Eq. (\ref{lambdalimit}) in the limit $m \gg m_P'$ and by Eq. (\ref{lambdalimit2}) in the limit $m \ll m_P'$, but distinct.
\\
\indent
For $m \geq m_P'$, the two solution branches $\lambda_{crit}^{\pm}$ explain the existence of the Schwarzschild radius as a quantum mechanical phenomenon due to the self-gravity of the wave packet, as well as the Compton radius. The latter is due to the usual quantum mechanical effects, which are now modified (even for small mass particles) due to self-gravity effects. Thus $\lambda_{crit}^{-}$, which is sub-Planckian even if physical observables such as $\Delta_{\psi}x$ remain super-Planckian, can be interpreted as giving rise to a Compton-type minimum radius, which still prevents the collapse of matter inside the black hole to a singularity, whereas $\lambda_{crit}^{+}$ gives rise to the Schwarzschild radius. In general $(\Delta_{\psi}x)_{min} \leq \Delta_{\psi}x \leq (\Delta_{\psi}x)_{max}$, where $2\pi(\Delta_{\psi}x)_{min} \approx 2\pi l_P$ and $2\pi(\Delta_{\psi}x)_{max} \approx \lambda_S$, asymptotically for $m \gg m_P'$. Interestingly, although the extended de Broglie relations allow sub-Planckian matter wave modes, the fact that $\Delta_{\psi}x \geq l_P$ for all $m$ implies that physical observables are limited to the super-Planckian regime. For $m \leq m_P'$, much the same is true, except that the limits on the localization scale (interpreted as the particle radius) give $2\pi(\Delta_{\psi}x)_{min} \geq \lambda_C \geq 2\pi l_P$ for $m \ll m_P'$. Again, although sub-Planckian matter waves are permitted, physical observables are limited to super-Planckian scales. 
\\
\indent
The extended de Broglie theory therefore provides a concrete manifestation of a minimum length uncertainty relation (MLUR). It has previously been suggested, using a variety of different arguments, that such relations should arise due to quantum gravitational effects (see, for example \cite{Chang:2011jj,Benczik:2002tt,Benczik:2002px}). More generally, it has been argued that quantum gravity implies the existence of a minimum length, usually assumed to be of the order of the Planck length (see \cite{Garay:1994en,Hossenfelder:2012jw} for reviews). The extended de Broglie theory is therefore consistent with such generic considerations and with some (but not \emph{all}) forms of MLUR already proposed in the literature. Specifically, the theory is compatible with the phenomenology implied by certain forms of the GUP, as mentioned previously, and discussed further below. 
\\
\indent
Interestingly, if recent experimental proposals to test the time-energy counterparts of position-momentum MLUR's using M{\" o}ssbauer neutrinos \cite{Raghavan:2012sy} are implemented, it may be possible to test relations of the form proposed here in the near future. It has also been suggested that such modified relations may be able to explain long-standing fundamental problems, such as the problem of quantum measurement/decoherence, CPT violation and the emergence of matter-antimatter asymmetry \cite{AmelinoCamelia:1997em}. 
\\
\indent
We may conjecture that a line defined by $\lambda_{S/C}'' = l_P^2/\lambda_{C/S}'$ (i.e. the reflection of the modified Compton-Schwarzschild line in $\lambda = l_P$) could be interpreted as giving the modified Schwarzschild and Compton radii in the sub-Planckian regime. Note the interchange of the indices $C$ and $S$, vis-a-vis the line $\lambda_{C/S}'$. Although not directly observable, a Compton-type limit of this form, extending into the region $m \geq m_P'$, may prevent singularity formation at the centre of a black hole. It would also be interesting to study its implications in the region $m \leq m_P'$ $-$ namely, the existence of a `quantum' Schwarzschild radius \emph{within} a fundamental particle.  
\\
\indent
Let us now compare these results with the predictions of the Generalized Uncertainty Principle (GUP). As one approaches the Planck energy from below, it has been argued~\cite{Adler_1, Adler_2, Adler_3, Adler_4} 
that the Heisenberg Uncertainty Principle (HUP), $\Delta x \Delta p \geq \hbar/2$, should be replaced by a GUP of the form
\begin{equation}
\Delta x > \frac{ \hbar}{\Delta p }+ \left( \frac{ \tilde{\alpha} l_P^2 }{ \hbar} \right) \Delta p \, .
\label{GUP1} 
\end{equation}
Here $\tilde{\alpha}$ is a dimensionless constant which depends on the particular model and the factor of $2$ in the first term has been dropped. We also use the notation $\Delta x$ and $\Delta p$ instead of $\Delta_{\psi} x$ and $\Delta_{\psi} p$ in the GUP, to indicate that we refer specifically to modifications of the HUP, rather than modifications of Eq. (\ref{SUP}). Eq. (\ref{SUP}), with $\hat{O}_1 = \hat{x}$ and $\hat{O}_2 = \hat{p}$ is also known as the Schr{\" o}dinger-Robertson Uncertainty Principle (SUP) \cite{Ro29,Sc30}, and is well defined in terms of  the fundamental state space structure of canonical quantum mechanics. By contrast, the `traditional' HUP is based on heuristic arguments and must be viewed as a postulate, with no rigorous foundation in the underlying mathematical structure of quantum theory \cite{Oz03A,Oz03B}. Variants of (\ref{GUP1}) can be found in other approaches to quantum gravity, including loop quantum gravity~\cite{ashtekar_1,ashtekar_2}, string theory~\cite{veneziano_1, veneziano_2, veneziano_3, veneziano_4, veneziano_5, veneziano_6}, non-commutative quantum mechanics \cite{majid}, gravitational ultraviolet self-completeness \cite{nicolini} and
general minimum length considerations \cite{maggiore_1,maggiore_2,maggiore_3}.
\\
\indent
If we rewrite (\ref{GUP1}) using the substitution $\Delta x \rightarrow \lambda$ and $\Delta p \rightarrow m c$, it becomes
\begin{equation}
\lambda > \tilde{\lambda}_C \equiv \frac{ \hbar}{mc} +\frac { \tilde{\alpha} Gm}{c^2} = \frac{\hbar}{mc} \left[ 1  + \tilde{\alpha}  \left( \frac {m}{m_P}\right)^2 \right] \, .
\label{GUP2}
\end{equation}
This expression may be regarded as a generalized Compton wavelength obtained from the GUP, the last term representing a small correction as one approaches the Planck point from below. However, one can also apply (\ref{GUP2}) for $m \gg m_P$ and it is interesting that in this regime it asymptotes to the Schwarzschild form, apart from a numerical factor. This suggests that  there is a different kind of positional uncertainty for an object larger than the Planck mass, related to the existence of black holes. This is not unreasonable since the Compton wavelength is below the Planck scale here. Also an outside observer cannot localize an object on a scale smaller than its Schwarzschild radius. 
\\
\indent
The GUP also has important implications for the black hole horizon size, as can be seen by  examining what happens as one approaches the intersect point from the right. In this limit, 
it is natural to write (\ref{GUP2}) as
 \begin{equation}
\lambda > \tilde{\lambda}_S =
 \frac{\tilde{\alpha} Gm}{c^2} \left[ 1  + \frac{1}{ \tilde{\alpha}} \left( \frac {m_P}{m}\right) ^2 \right] 
\label{GEH2}
\end{equation}
and this represents a small perturbation to the Schwarzschild radius for $m \gg m_P$ if one assumes $\alpha =2$. There is no reason for anticipating $\alpha=2$ in the heuristic derivation of the GUP. However, the factor of 2 in the expression for the Schwarzschild radius is precise, whereas  the coefficient associated with the Compton term is somewhat arbitrary. This motivates an alternative approach in which the free constant in (\ref{GUP2}) is associated with the first term rather than the second. One then replaces Eqs.~(\ref{GUP2}) and (\ref{GEH2}) with the expressions
 \begin{equation}
\tilde{\lambda}_C = 
\frac{\tilde{\beta} \hbar}{mc} \left[ 1  + \frac{2}{\tilde{\beta}} \left( \frac {m}{m_P}\right)^2 \right] 
\label{GUP4A}
 \end{equation}
and 
\begin{equation}
\tilde{\lambda}_S = \frac{2 Gm}{c^2} \left[ 1  + \frac{\tilde{\beta}}{2} \left( \frac {m_P}{m}\right) ^2 \right] 
\label{GUP4B}
\end{equation}
for some constant $\tilde{\beta}$, with
the second expression being regarded as a Generalized Event Horizon (GEH). In principle, the constants $\tilde{\alpha}$ in Eq.~\eqref{GUP2} and $\tilde{\beta}$ in Eq.~\eqref{GUP4B} could be independent.
\\
\indent
These arguments suggest that there is a connection between the Uncertainty Principle on microscopic scales and black holes on macroscopic scales. This is termed the Black Hole Uncertainty Principle (BHUP) correspondence and it is manifested in a unified expression for the Compton wavelength and Schwarzschild radius \cite{Ca:2014}. It is a natural consequence of combining the notions of the GUP and the GEH. Indeed, it would be satisfied for any form of the function $\tilde{\lambda}_C \equiv \tilde{\lambda}_S$ which asymptotes to $\lambda_C$ for $m \ll m_P$ and $\lambda_S$ for $m \gg m_P$. Models in which this function is symmetric under the duality transformation $m \leftrightarrow const. \times m_P^2/m$ are said to satisfy the {\it strong} BHUP correspondence \cite{Ca:2014}.
\\
\indent
Recently, an alternative interpretation of Eq.~\eqref{GUP4B} has been suggested \cite{cmn} (henceforth CMN), in which  one distinguishes between $m$ and the ADM mass
\begin{equation}
m_{ADM} = m \left(1+\frac{\tilde{\beta}}{2} \frac{m_P^2}{m^2}\right) \, .
\label{newmass}
\end{equation}
This is close to $m$ for $m \gg m_P$ but it reaches  a minimum value of $\sqrt{\tilde{\beta}/2} \, m_P$ as $m$ decreases and then diverges as $m \rightarrow 0$. Clearly $m_{ADM}$ is closely related to the quantity $\mu$ defined by Eq.~\eqref{mu}. Indeed, one can make the identification $m_{ADM} = \mu m_P'$ providing $\tilde{\beta} = \pi / \sqrt{2}$. This version of the GUP therefore seems particularly close to the present approach, though one fundamental difference remains: in standard GUP scenarios, the Schwarzschild remains a \emph{minimum} radius, rather than a maximum. 
%
%Section5%%%%%%%%%%%%%%%%%%%%%%%%%%%%%%%%%%%%%%%%%%%%%%%%%%%%%%%%%%%%%%%%``
%%%%%%%%%%%%%%%%%%%%%%%%%%%%%%%%%%%%%%%%%%%%%%%%%%%%%%%%%%%%%%%%%%%%%
\section{Extended de Broglie relations and the Hawking temperature} \label{Sec.5}
The Hawking temperature for black holes may be derived heuristically in canonical non-relativistic quantum mechanics by assuming `saturation' of the UP in the following sense. Keeping in mind the direction of the inequality implied by the heuristic UP-based argument, one has an upper limit  on $\Delta_{\psi}x$ associated with the size of the black hole and a corresponding lower limit on $\Delta_{\psi}p$:
\begin{eqnarray}\label{}
(\Delta_{\psi}x)_{max} \approx \lambda_S = \frac{2ml_P}{m_P}, \ \ \ (\Delta_{\psi}p)_{min} \approx \frac{1}{2}
\frac{\hbar}{\lambda_S} =
 \frac{1}{4} \frac{m_P^2c}{m}.
\end{eqnarray}
A temperature $T_H$ may then be defined by
\begin{eqnarray}\label{}
(\Delta_{\psi}p)_{min} = \chi m_Pc \frac{T_H}{T_P}\, ,
\end{eqnarray}
where $T_P = m_Pc^2/k_B$ is the Planck temperature and $\chi$ is a constant. Agreement with the Hawking temperature requires $\chi = 2 \pi$, so that
\begin{eqnarray}\label{T_H}
T_H = \frac{\hbar c^3}{8\pi Gm k_B} = 
\frac{1}{8\pi}\frac{m_P T_P}{ m} \, .
\end{eqnarray}
As suggested in \cite{LaCa:2015}, we may also invoke the saturation condition to define a lower limit on the size of a particle and an associated upper limit on the momentum uncertainty:
\begin{eqnarray}\label{}
(\Delta_{\psi}x)_{min} \approx \lambda_C = \pi \frac{m_P'l_P}{m}, \ \ \ (\Delta_{\psi}p)_{max} \approx \frac{1}{2}\frac{\hbar}{\lambda_C} = \frac{1}{\pi^2}mc \, .
\end{eqnarray}
This defines the Compton line, with the numerical factor chosen to ensure continuity at its intersect with the Schwarzschild line at $m = m_P'$. We then define a Compton temperature by
\begin{eqnarray}\label{T_C}
T_C = \frac{1}{\chi \pi^2}\frac{m}{m_P}T_P = \frac{1}{2 \pi^3}\frac{mc^2}{k_B} \, ,
\end{eqnarray}
where $\chi = 2 \pi$ again ensures that $T_C$ and $T_H$ meet at  $m = m_P'$. In this case, the direction of the inequality implied by the UP changes at $m = m_P'$, so that the Hawking temperature may be regarded as a {\it minimum} temperature and the Compton temperature as a {\it maximum} temperature. This gives $T_C = T_H = (2\pi)^{-2}T_P$ at the critical point and the two temperatures are plotted, as red and blue lines, respectively, as functions of $m$ in Fig. 8. 
\\
\indent
In the extended de Broglie relations theory, we have
\begin{eqnarray}\label{}
2\pi\Delta_{\psi}x &\equiv& \lambda_{C/S}' =2\pi [\lambda_{crit}^{\pm}/(2\pi) + (\pi/2)l_P^2/\lambda_{crit}^{\pm}] \, , 
\nonumber\\
(\pi/2)\Delta_{\psi}p &\equiv& \frac{\pi}{2}\frac{\hbar}{\lambda_{C/S}'} = \frac{\pi}{4}\frac{l_Pm_Pc}{[\lambda_{crit}^{\pm}/(2\pi) + (\pi/2)l_P^2/\lambda_{crit}^{\pm}]} = 2\pi m_Pc \frac{T_{C/H}'}{T_P} \, ,
\label{Tprime}
\end{eqnarray}
where, by our previous convention, we use $\lambda_{C}'$ and $T_{C}'$ for $m \leq m_P'$ and $\lambda_{S}'$ and $T_{H}'$ for $m \geq m_P'$. In this notation, $T_{C}'$ represents the generalized Compton temperature, which differs from Eq.~(\ref{T_C}) due to the self-gravity of the wave packet, and $T_{H}'$ represents the generalized Hawking temperature, which is likewise modified due to quantum gravitational effects. Then $T_C' = T_H' = (16\pi)^{-1}T_P$ at $m = m_P'$ but $T_C' \approx T_C = (2 \pi^3)^{-1}(mc^2/k_B)$ for $m \ll m_P'$ and  $T_H' \approx T_H = (8 \pi)^{-1}(\hbar c^3/Gm k_B)$ for $m \gg m_P'$, as required. The $T_{C/H}'$ curves are also shown (in green), together with the $T_{C}$ and $T_{H}$ asymptotes in Fig. 8.
%
%Fig. 8
\begin{figure}[h] \label{Fig.8}
\centering
\includegraphics[width=12cm]{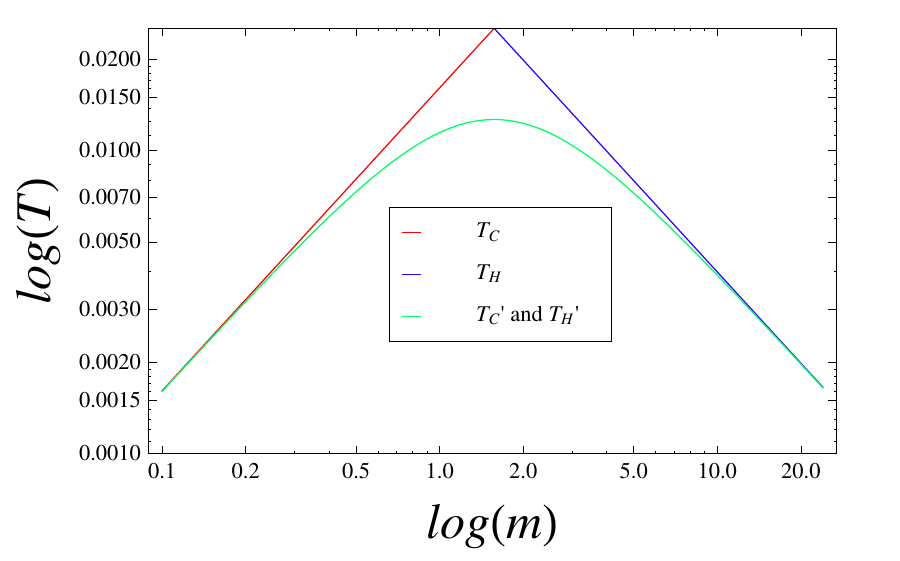}
\caption{Comparing modified Compton and Hawking temperatures, $T_C'$ and $T_H'$, in the extended de Broglie relations theory, with the Compton temperature $T_C$ in canonical quantum theory and the Hawking temperature $T_H$ in semi-classical gravity.}
\end{figure}
\\ 
\indent
We now recall the determination of the temperature in the usual GUP approach \cite{cmn}. Adler {\it et al.}  assume that  $\Delta x$ is associated  with the Schwarzschild radius but that $\Delta p$ and $\Delta x$ are related by the linear form of the GUP. In this case, identifying the size of the black hole with the wavelength of the emitted radiation gives
\begin{equation}
\frac{2Gm}{c^2} = \frac{\hbar \eta c}{k_BT} + \frac {\tilde{\alpha} G k_B T}{\eta c^4}
\end{equation}
where $\eta$ is a numerical constant; this is equivalent to $1/(2 \chi)$ where $\chi$ is the constant appearing in the extended de Broglie analysis. This leads to a temperature 
\begin{equation}
T = {\eta m c^2 \over \tilde{\alpha} k_B} \left(1- \sqrt{1- \frac{\tilde{\alpha} m_P^2}{m^2}} \right) \, ,
\label{adlertemp2}
\end{equation}
giving a small perturbation to the standard Hawking temperature,
\begin{equation}
T \approx
{\eta \hbar c^3 \over 2 G k m}  \left[ 1 + {\tilde{\alpha} m_P^2 \over 4m^2} \right] \, ,
\end{equation}
for $m \gg m_P$, if we choose $\eta = (4\pi)^{-1}$. However, the exact expression becomes complex when $m$ falls below $\sqrt{\tilde{\alpha}} \, m_P$, corresponding to a minimum mass and a maximum temperature.  
\\
\indent
We obtain a different result if we express $T$ in terms of $m$ by identifying the size of the GEH (rather than the Schwarzschild radius) with the wavelength of the emitted radiation:
\begin{eqnarray}
 \left( \frac {\hbar \eta c}{k_BT} \right) + \left(\frac{\tilde{\alpha} l_P^2 k_BT}{\hbar \eta c} \right)
= \left( \frac{\hbar \tilde{\beta}}{mc} \right) + \left( \frac{2Gm}{c^2}\right) \, .
\end{eqnarray}
This implies
\begin{eqnarray}
k_BT = \frac{\eta m c^2}{\tilde{\alpha}} f(m, \tilde{\alpha}, \tilde{\beta})
\label{rainbowtem}
\end{eqnarray}
where the function $f$ is real for $\tilde{\alpha} < 2 \tilde{\beta}$ and given by 
\begin{eqnarray}
f(m, \tilde{\alpha}, \tilde{\beta}) = 
1 + {\tilde{\beta} \over 2} \left({m_P \over m}\right)^{2} 
\pm  \sqrt{1+ (\tilde{\beta} - \tilde{\alpha})\left( {m_P \over m} \right)^{2} + {\tilde{\beta}^2 \over 4} \left({m_P \over m}\right)^{4} } \, .
\end{eqnarray}
This can be approximated by
\begin{equation}
T \approx {\eta \hbar c^3 \over 2 G k_B m} \left[ 1 - \left( \frac {2\tilde{\beta} - \tilde{\alpha}}{4}\right)  \left( {m_P \over m} \right)^{2} \right]  
\label{super}
\end{equation}
for $m \gg m_P$ and
\begin{equation}
T \approx {\eta m c^2 \over k_B \tilde{\beta}} \left[1 - \left(\frac{2\tilde{\beta} - \tilde{\alpha}}{\tilde{\beta}^2}\right)  \left({m \over m_P} \right)^{2} \right]   
\label{sub}
\end{equation}
for $m \ll m_P$. The special case $\tilde{\alpha} = 2 \tilde{\beta}$, associated with the BHUP correspondence, gives
\begin{eqnarray}
k_BT = \mathrm{min} \left[\frac{\hbar \eta c^3}{2Gm} \,  , \, \frac{2\eta mc^2}{\tilde{\alpha}} \right]  \, .
\label{mess3}
\end{eqnarray}
The first expression is the {\it exact} Hawking temperature, with no small correction term, but one must cross over to the second expression below $m = \sqrt{\tilde{\alpha} /4} \, m_P$ in order to avoid the temperature going above the Planck value $T_P = m_P c^2/k_B$. 
\\
\indent
In the CMN model \cite{cmn}, the temperature is still given  by (\ref{adlertemp2}) providing $m$ is interpreted as the ADM mass~\eqref{newmass}. Since this has a minimum of $\sqrt{\tilde{\beta}/2} \, m_P$, the temperature reaches a maximum and then decreases  for $\tilde{\alpha} < 2 \tilde{\beta}$ rather than going complex. We note that temperature \eqref{adlertemp2} is qualitatively similar to $T_{C/H}'$ given by Eq.~\eqref{Tprime}. However, CMN also calculate the black hole temperature using the surface gravity argument and this gives a slightly different result: 
\begin{eqnarray}
T = \frac{m_P^2}{8\pi m(1+\tilde{\beta} m_P^2/2m^2)} \frac{c^2}{k_B} \, .
\label{hawktemp}
\end{eqnarray}
The limiting behaviour in the different mass regimes is as follows: 
\begin{eqnarray}
m \gg m_P & \Longrightarrow & T \approx \frac{m_P^2}{8\pi m} \left[ 1- \tilde{\beta}  \left( {m_P \over m} \right)^{2} \right] \frac{c^2}{k_B} \label{bigm}\\
m \approx m_P & \Longrightarrow & T \approx \frac{m_P}{8\pi(1+\tilde{\beta}/2)} \frac{c^2}{k_B} \label{mplm}\\
m \ll m_P & \Longrightarrow & T \approx \frac{m}{4\pi \tilde{\beta}} \left[ 1- \frac{1}{\tilde{\beta}}  \left( {m \over m_P} \right)^{2} \right] \frac{c^2}{k_B} \, . \label{smallm}
\end{eqnarray}
The large $m$ limit (\ref{bigm}) is the usual Schwarzschild temperature but, as the black hole evaporates, it reaches a maximum temperature (\ref{mplm}). The temperatures given by (\ref{rainbowtem}) and (\ref{hawktemp}) agree to first order but not to second order. 

The point of this discussion is to emphasize that there are several possible forms for the mass dependence of the black hole temperature, depending on the precise form of the GUP, and none of these is exactly the form predicted by our extended de Broglie relations. However, the latter derive from Eqs. (\ref{deBroglie_all_m})-(\ref{kappa1}), which merely represent the simplest possible situation and are only applicable in the non-relativistic context anyway. Therefore it is premature to conclude which prediction is correct. The important point is that all the expressions agree asymptotically, tending to $T_H$ for $m \gg m_P$ and $T_C$ for $m \ll m_P$.Ó
%
%Section6%%%%%%%%%%%%%%%%%%%%%%%%%%%%%%%%%%%%%%%%%%%%%%%%%%%%%%%%%%%%%%%%``
%%%%%%%%%%%%%%%%%%%%%%%%%%%%%%%%%%%%%%%%%%%%%%%%%%%%%%%%%%%%%%%%%%%%%
\section{Discussion} \label{Sec.6}
We have proposed a modification of canonical quantum theory based on modified de Broglie relations, which allow the concept of a matter wave to be extended to `particles' with energies above the Planck scale (i.e. black holes). The extended de Broglie relations give GUP-type phenomenology, leading to the Compton-Schwarzschild correspondence, also known as the Black Hole Uncertainty Principle (BHUP) correspondence \cite{Ca:2014}, without violating the fundamental UP derived from the state space structure of quantum mechanics. This state space structure is crucial to both the mathematical self-consistency of the theory and its physical interpretation as a probabilistic theory. Any phenomenological modifications to the standard UP, motivated by such heuristic considerations as the disturbance to the quantum mechanical system caused by the gravitational field of the probing particle, must be checked for mathematical consistency. This may be possible using theories of deformed quantum mechanics, in which the deformation parameter is explicitly linked to a change in the background metric \cite{Pe11,Maziashvili:2011dx,Faizal:2015kqa}. However, it remains unclear whether \emph{any} GUP-type phenomenology which leads to a unified expression for the Compton and Schwarzschild scales can account for their different physical natures. In particular, the Compton radius  represents the minimum scale within which the mass of a quantum mechanical particle is localized, whereas the Schwarzschild radius represents the maximum localization radius. One advantage of the extended de Broglie theory is that it yields a minimum localization scale for $m \leq m_P'$ and a maximum localization scale for $m \geq m_P'$ (where $m_P' = (\pi/2)m_P \approx m_P$), as required. 
\\
\indent
One drawback of the non-relativistic analysis is that our derivations of the modified Compton and Schwarzschild expressions, $\lambda_C'$ and $\lambda_S'$, do not really apply in the relativistic context. This is analogous to the fact that the standard Compton wavelength expression, $\lambda_C$, does not apply in canonical non-relativistic quantum theory. Rather, it is a relativistic result, derived heuristically from the non-relativistic theory, that marks the point at which the non-relativistic theory itself becomes invalid. Further work is required to determine whether the results contained in the present work can be reproduced in a relativistic model.  
\\
\indent
It is encouraging that  expressions for the Schwarzschild and Compton scales may be obtained in a n{\" a}ive non-relativistic theory by introducing extended de Broglie relations for energies $E > m_Pc^2$. We therefore consider the possibility that combining extended relations with fully relativistic energy-momentum relations may yield relativistic field theories (or at least theories in which Lorentz invariance is violated only close to the Planck scale) which correctly incorporate quantum gravitational effects and key features of classical gravity in the appropriate limit. Nonetheless, precise calculations are required to establish concrete predictions in the relativistic-gravitational regime. 
\\
\indent
In this context, the issue arises of whether local Lorentz invariance applies when moving to a fully relativistic theory with extended de Broglie relations. Specifically, we must consider the questions: Is Lorentz invariance violated on scales close to $l_P$ or $t_P$ and, if so, how does the extended theory relate to models that violate Lorentz invariance close to the Planck scales, such as doubly special relativity \cite{AmelinoCamelia:2002wr} or gravity's rainbow \cite{rainbow}? Does making extended relations consistent with the relativistic energy-momentum relation fix the form of the dispersion relations, or are multiple formulations which give the same asymptotic behavior possible? For example, do we need to add terms in quadrature in order to obtain a self-consistent generalization of the Dirac or Klein-Gordon equations? Extension to the fully relativistic case is clearly important, but conceptual and technical difficulties resulting from such questions remain.
\\
\indent
An advantage of the extended de Broglie theory is that it provides a unifed ontology, in which the Compton and Schwarzschild expressions in all four sectors of the $(\log \ m,\log \ \lambda)$ diagram arise from the same underlying theory. We no longer need to extrapolate classical gravity theory (either Newtonian gravity with a maximum velocity $v_{max}=c$ or general relativity) into regions where quantum effects become important. Nor need we extrapolate non-gravitational quantum theory into regimes in which gravitational effects become significant. The extended de Broglie relations yield features of classical gravitation in the quantum regime and features of canonical quantum mechanics in the gravitational regime, which bolsters our claim that the extended relations should be interpreted as representing quantum gravity effects. Since sub-Planckian solutions are a formal mathematical feature of the extended relations, we should take seriously the possibility of extending the notion of de Broglie matter waves into the sub-Planckian region, even though it is unclear how to interpret them physically. It is encouraging that the results of this study suggest that, even if we allow sub-Planckian modes, physical observables such as $\Delta_{\psi}x$ or particle/black hole radii are still super-Planckian. 
\\
\indent
In extended de Broglie theory, the Hamiltonian and momentum operators are defined via expansions which contain an infinite number of terms, so we formally require an infinite number of boundary conditions to solve the quantum EOM. In classical physics, a theory requiring such conditions would be considered unphysical. However, even in canonical quantum physics, the initial conditions are fixed up to any order, since the EOM must act on an initial (known) wave function $\psi$. The same applies here, so in principle this is no problem. One crucial difference between the extended and canonical theories is that we cannot predict the quantum evolution of $\psi$ with \emph{complete} precision. Nonetheless, we \emph{can} predict it with arbitrary precision, by taking enough terms in the series to compare with experimental data. 
\\
\indent
As with any quantum theory, a further complication arises because the quantum EOM depend on the form in which the dispersion relations are written. The EOM corresponding to a specific dispersion relation is unique in classical mechanics but not in quantum theory. The most famous example is the difference between the Klein-Gordon equation, which corresponds to substituting the de Broglie relations into $E^2 = p^2c^2 + m^2c^4$, and the Dirac equation, which corresponds to substituting them into $E = \pm\sqrt{p^2c^2 + m^2c^4}$. In the extended de Broglie theory, we face an additional ambiguity in that, not only the form of the classical energy-momentum relation matters, but also how we rearrange it \emph{after} substitution of the quantum relations. Our definitions are motivated by the need to consistently interpret the EOM as a quantum analogue of a classical energy-momentum relation in which the operators corresponding to energy and momentum can be clearly identified (and computed). However, this prescription is not unique and other (inequivalent) quantum theories may exist, even for the same relations. Whether or not this ambiguity is desirable for developing a consistent theory of quantum gravitational interactions is debatable. In principle, each new theory may describe the interactions of different kinds of fundamental particle which couple to gravity in different ways, just as the inequivalent Klein-Gordon and Dirac theories describe the quantum dynamics of bosons and fermions.
\\
\indent
Finally, we note that it would also be interesting to consider the implications of extended de Broglie theory in the context of higher dimensions. The Schwarzschild radius of a black hole is independent of Planck's constant \emph{only} in $d=3$ spatial dimensions, so that higher-dimensional black holes are necessarily both gravitational and quantum objects. Though the standard Compton wavelength is independent of Newton's constant, an argument for the dimensional dependence of the Compton formula is proposed in \cite{LaCa:2015} which, if correct, implies that fundamental particles are manifestly quantum \emph{and} gravitational for $d \neq 3$. This argument is proposed within the framework of canonical quantum theory, but it would be interesting to apply analogous reasoning in the context of the extended relations presented here. In this case, both black holes and fundamental particles are manifestly quantum and gravitational even in three spatial dimensions. 
%
%%%%%%%%%%%%%%%%%%%%%%%%%%%%%%%%%%%%%%%%%%%%%%%%%%
%%%%%%%%%%%%%%%%%%%%%%%%%%%%%%%%%%%%%%%%%%%%%%%%%%``
\renewcommand{\theequation}{A-\arabic{equation}}
\setcounter{equation}{0}  % reset counter 
\section*{A \ Unitary evolution of self-gravitating states} \label{Appendix A}
In this section we demonstrate that the extended de Broglie relations satisfy unitarity in all four regimes. In canonical quantum mechanics, the Schr{\" o}dinger equation with \emph{general} time-independent Hamiltonian $\hat{H}(x)$ is 
\begin{eqnarray} \label{Schrod_QM}
i\hbar\frac{\partial \psi}{\partial t} = \hat{H}(x)\psi \, .
\end{eqnarray} 
It follows that the time evolution of a single momentum eigenstate $\phi(x,0) = e^{ikx}$, existing at $t=0$, is given by the unitary operator
\begin{eqnarray} \label{U_k_QM}
\hat{U}_k(t) = e^{-i\omega(k)t}\mathbb{I} = e^{-iE_kt/\hbar}\mathbb{I} \, ,
\end{eqnarray}
where $\hat{H}(x)e^{ikx} = E_ke^{ikx}$ and $E_k = \hbar\omega(k)$. This ensures that the first  de Broglie relation, $E = \hbar\omega$, holds for each individual eigenfunction $\phi(x,t) = e^{i(kx - \omega(k)t)}$, corresponding to the momentum eigenvalue $k$. However, the form of $\omega(k)$ cannot be determined without specifying the Hamiltonian.

For general wave packets,
\begin{eqnarray} \label{wavepacket_QM}
\psi(x,t) = \int_{-\infty}^{\infty}a(k)e^{i(kx-\omega(k)t)}dk \, ,
\end{eqnarray}
Eq.~(\ref{Schrod_QM}) yields 
\begin{eqnarray} \label{wavepacket_subs_QM}
\hbar\int_{-\infty}^{\infty}a(k)\omega(k)e^{i(kx-\omega(k)t)}dk = \int_{-\infty}^{\infty}a(k)\hat{H}(x)e^{i(kx-\omega(k)t)}dk \, ,
\end{eqnarray}
which holds if and only if $\hat{H}(x)e^{ikx} = \hbar\omega(k)e^{ikx}$ for all $k$. It follows from the definition of a function of an operator \cite{Ish95} that the time evolution of a general initial wave packet,  $\psi (x,0)  = \int_{-\infty}^{\infty}a(k)e^{ikx}dk$, is given by 
\begin{eqnarray} \label{U_QM}
\hat{U}(x,t) = e^{-i\hat{H}(x)t/\hbar}\mathbb{I} \, .
\end{eqnarray}
For \emph{free} particles, we therefore require $\hat{H}(x) = -\hbar^2/(2m)(\partial^2/\partial x^2)$. This ensures that the second de Broglie relation, $p = \hbar k$, holds in conjunction with the non-relativistic energy-momentum relation, $E = p^2/(2m)$, giving $\omega(k) = \hbar k^2/(2m)$. 

We now consider the time evolution operator corresponding to Eq.~(\ref{H:m<m_P:w<w_p:k<k_P}), which we first rewrite as 
\begin{eqnarray} \label{H:m<m_P:w<w_p:k<k_P-Appendix}
i\hbar [ 1 - \omega_P^{-2}(\partial^2 /\partial t^2) ]^{-1}\frac{\partial \psi}{\partial t} = \hat{H}(x)\psi \, ,
\end{eqnarray} 
where
\begin{eqnarray} \label{H:m<m_P:w<w_p:k<k_P_H(x)}
\hat{H}(x) = -\frac{\hbar^2}{2m}[1 - k_P^{-2}(\partial^2 /\partial x^2)]^{-2}\frac{\partial^2 }{\partial x^2} \, .
\end{eqnarray} 
Recall that both $\hat{H}(x)$ in (\ref{H:m<m_P:w<w_p:k<k_P_H(x)}) and the terms inside the square brackets in (\ref{H:m<m_P:w<w_p:k<k_P-Appendix}) are defined via the expansions in Eqs.~(\ref{Exp1A}) and (\ref{Exp1B}), respectively, when acting on momentum eigenstates of the form $\phi(x,t) = e^{i(kx - \omega(k)t)}$.  The latter implies a \emph{pair} of time evolution operators for the single momentum eigenstate $\phi(x,0) = e^{ikx}$: 
\begin{eqnarray} \label{U_k_1}
\hat{U}_k^{\pm}(t) = \exp\left[{-i\omega_{\pm}(k)t}\right]\mathbb{I} = \exp\left[-\frac{i\hbar \omega_P^2}{2E_k}\left(1 \pm \sqrt{1-\frac{4E_k^2}{\hbar^2 \omega_P^2}}\right)t\right]\mathbb{I} \, 
\end{eqnarray}
with $E_k \equiv E_{\omega(k)} = \hbar\omega_P^2[\omega(k) + \omega_P^2/\omega(k)]^{-1}$. Note that Eq.~(\ref{U_k_1}) holds for \emph{any} Hamiltonian yielding $\hat{H}(x)e^{ikx} = E_ke^{ikx}$, not just the one given in (\ref{H:m<m_P:w<w_p:k<k_P_H(x)}). If, in addition, we invoke the definition of the Hamiltonian given by Eqs.~(\ref{H:m<m_P:w<w_p:k<k_P_H(x)}) and (\ref{Exp1A}), we obtain 
\begin{eqnarray}
E_k = \frac{\hbar^2k_P^4}{2m}\left(k +\frac{ k_P^2}{k}\right)^{-2} \, .
\end{eqnarray}

Equation~(\ref{U_k_1}) shows that, when considering single modes, a subtlety arises  in the extended de Broglie theory that is not present in canonical quantum mechanics: two solution branches $\omega_{\pm}(k)$ exist for a given value of $k$. However, this does not matter when considering a single quadrant of the $(k,\omega(k))$ diagram, since the solutions are related via $\omega_{\pm}(k) = \omega_P^2/\omega_{\mp}(k)$. For example, in Sec. \ref{Sec.3.3} we conjectured that Eq. (\ref{H:m<m_P:w<w_p:k<k_P}) is valid for modes in the range $\omega < \omega_P$ and $k < k_P$, corresponding to particles with $m < m_P'$, since \emph{only} in this range are the expansions (\ref{Exp1A})-(\ref{Exp1B}) well defined for each eigenfunction $\phi(x,t) = e^{i(kx - \omega(k)t)}$. Therefore, when considering single modes belonging to the  $\omega_{-}(k)$ branch, governed by Eq. (\ref{H:m<m_P:w<w_p:k<k_P}), we may neglect modes in the $\omega_{+}(k)$ branch, governed by (\ref{H:m<m_P:w>w_p:k>k_P})/(\ref{H:m<m_P:w<w_p:k<k_P-Appendix}), and consider only $\hat{U}_k^{-}(t) $.

Nonetheless, this raises the interesting question of what happens if wave packets are composed of modes with both super-Planckian and sub-Planckian wavelengths. Since these modes obey different equations of motion (EOM), they should evolve independently and can be interpreted as separate particles with the same mass but different physical properties. As such, in the extended de Broglie theory of \emph{free} particles, super-Planckian and sub-Planckian wavelengths do not mix but form separate non-interacting wave packets. The implications of this for the dark matter and dark energy paradigms, as well as the possibility of including some form of entanglement between super-Planckian and sub-Planckian particles, is left for a future paper. 

When dealing with wave packets, a second subtlety in the extended theory is that the integral over $dk$ in Eq. (\ref{wavepacket_QM}) is not well defined everywhere in the range $k \in (-\infty,\infty)$, even \emph{within} a single solution branch $\omega_{\pm}(k)$, due to the existence of maximum and minimum wave numbers for a given particles mass, $k_{\pm}(m)$. The reality of both solutions requires either
\begin{eqnarray} \label{k^2>k_+^2}
k^2 \geq k_{+}^2 \iff k > k_{+}, \ k < -k_{+}
\end{eqnarray}
or
\begin{eqnarray} \label{k^2<k_-^2}
k^2 \leq k_{-}^2 \iff k < k_{-}, \ k > -k_{-}
\end{eqnarray}
but one range may be discarded for a single quadrant of the $(k,\omega(k))$ diagram, governed by a single EOM, since the limits are related by $k_{\pm}(m) = k_P^2/k_{\mp}(m)$. We therefore replace the expression in Eq. (\ref{wavepacket_QM}) with 
\begin{eqnarray} \label{wavepacket_ext}
\psi(x,t) = \int_\Sigma a(k)e^{i(kx-\omega(k)t)}dk \, ,
\end{eqnarray}
where $\Sigma \subset \mathbb{R}$ is the integral range corresponding to the quadrant considered. 

For example, for wave packets governed by Eq.~(\ref{H:m<m_P:w<w_p:k<k_P-Appendix}), the integration must be performed over the range $\Sigma = [-k_{-}(m),k_{-}(m)]$, where $k_{-}(m)$ is given by Eq. (\ref{disp_rel_m<m_P4}). Substituting the wave packet (\ref{wavepacket_ext}) into (\ref{H:m<m_P:w<w_p:k<k_P-Appendix}) gives
\begin{eqnarray} \label{wavepacket_subs_1}
\hbar\omega_P^2\int_{-k_-}^{k_-} a(k)[\omega(k) + \omega_P^2/\omega(k)]^{-1}e^{i(kx-\omega(k)t)}dk = \int_{-k_-}^{k_-}
a(k)\hat{H}(x)e^{i(kx-\omega(k)t)}dk \, ,
\end{eqnarray}
which holds if and only if $\hat{H}(x)e^{ikx} = \hbar\omega_P^2[\omega(k) + \omega_P^2/\omega(k)]^{-1}e^{ikx}$ for all $k \in [-k_{-}(m),k_{-}(m)]$. From the definition of a function of an operator \cite{Ish95}, the time evolution of a general initial wave packet, $\psi(x,0) = \int_{\Sigma}a(k)e^{ikx}dk$, governed by Eq.~(\ref{H:m<m_P:w<w_p:k<k_P-Appendix}), is then given by 
\begin{eqnarray} \label{U_1}
\hat{U}^{-}_{m<m_P'}(x,t) = \exp\left[-\frac{i \hbar \omega_P^2}{2\hat{H}_{-}(x)}\left(1 - \sqrt{1-\frac{4\hat{H}_{-}^2(x)}{\hbar^2 \omega_P^2}}\right)t\right]\mathbb{I} \, .
\end{eqnarray}
Here $\hat{H}_{-}(x)$ denotes the Hamiltonian (\ref{H:m<m_P:w<w_p:k<k_P_H(x)}) for the negative solution branch, $\omega_{-}(k)$, for $m < m_P'$. Thus the evolution is \emph{non-canonical but unitary}. The operator (\ref{U_1}) gives the time evolution of a fundamental, self-gravitating, quantum mechanical particle which is `ordinary' in the sense that each mode in its wave packet expansion has a wavelength greater than the Planck length.

Using $\hat{H}(x)$ defined by Eqs.~(\ref{H:m<m_P:w<w_p:k<k_P_H(x)}) and (\ref{Exp1A}), if $E_k \ll (1/2)\hbar\omega_P = 2m_P'c^2$ for $k \ll k_P$, which is equivalent to taking the limit $m \ll m_P'$, we have $\omega_{-}(k) \approx E_k/\hbar \approx \hbar k^2/(2m)$. Thus the negative solution branch of Eq.~(\ref{U_k_1}) obeys the standard time evolution (\ref{U_k_QM}) for a single eigenstate. When these conditions hold for all $k \in [-k_{-}(m),k_{-}(m)]$, Eq.~(\ref{U_1}) gives the standard time evolution (\ref{U_QM}) for a wave packet. The wave packet expansion (\ref{wavepacket_ext}) also reduces to the standard form, Eq. (\ref{wavepacket_QM}), since $|k_{-}(m)| \rightarrow \infty$ in the limit $m/m_P' \rightarrow 0$. In other words, for $m \ll m_P'$, corrections due to gravitational effects become negligible and the extended de Broglie theory obeys the time evolution and expansion theorems of canonical quantum mechanics. 

Next we consider the time evolution operator implied by Eq.~(\ref{H:m>m_P:w<w_p:k<k_P}), which governs quantum `particles' with masses $m > m_P'$ (i.e. black holes) and again corresponds to wave packets in which each mode has a super-Planckian wavelength. Following our previous procedure, we first rewrite Eq. (\ref{H:m>m_P:w<w_p:k<k_P}) as
\begin{eqnarray}  \label{H:m>m_P:w<w_p:k<k_P-Appendix}
\frac{i\hbar}{8} [1 - k_P^{-2} (\partial^2 /\partial x^2)]^2\frac{\partial \psi}{\partial t} = \hat{H}'(x,t)\psi \, ,
\end{eqnarray} 
where
\begin{eqnarray}  \label{H:m>m_P:w<w_p:k<k_P_H(x)}
\hat{H}'(x,t) = -mc^2k_P^{-2}[1 -  \omega_P^{-2}(\partial^2 /\partial t^2)]\frac{\partial^2 }{\partial x^2} \, . 
\end{eqnarray}
Then the time evolution operator for a single momentum eigenstate, $\phi(x,0) = e^{ikx}$, is 
\begin{eqnarray} \label{U_k_2}
\hat{U}_{(k,\omega(k))}(t) = \exp\left[{-i\omega(k)t}\right]\mathbb{I} = \exp\left[-\frac{8i E_{(k,\omega(k))}}{\hbar}\left(1 + \frac{k^2}{k_P^2}\right)^{-2}t\right]\mathbb{I} \, 
\end{eqnarray}
with $E_{(k,\omega(k))} = (\hbar\omega(k)/8)(1 + k^2/k_P^2)^2$. Eq.~(\ref{U_k_2}) holds for \emph{any} Hamiltonian yielding an energy eigenvalue that depends on $k$ both explicitly and implicitly via $\omega(k)$, $\hat{H}(x,t)e^{ikx} = E_{(k,\omega(k))}e^{ikx}$. If, in addition, we invoke the definition of the Hamiltonian given by Eq.~(\ref{H:m>m_P:w<w_p:k<k_P_H(x)}), we have
\begin{eqnarray} \label{E_k_omega(k)}
E_{(k,\omega(k))} =  \frac{\hbar\omega(k)}{8}\left(1 + \frac{k^2}{k_P^2}\right)^2 = mc^2\frac{k^2}{k_P^2}\left(1+ \frac{\omega^2(k)}{\omega_P^2}\right).
\end{eqnarray}
From this, we obtain
\begin{eqnarray} \label{omega(k)_1}
\omega_{\pm}(k) = \frac{2E_k}{\hbar}\left(1 \pm \sqrt{1 - \frac{\hbar^2\omega_P^2}{4E_k^2}} \right)
\end{eqnarray}
with
%CORRECTED EQ. IN arXiv_v3
\begin{eqnarray} 
E_{(k,\omega(k))} = \frac{\hbar\omega(k)}{8}\left(1 + \frac{k^2}{k_P^2}\right)^2 
\equiv E_k = \frac{\hbar^2}{32m}\left(k+\frac{k_P^2}{k}\right)^2 \, . 
\end{eqnarray} 
%EQ. IN JHEP PUBLISHED VERSION
%\begin{eqnarray} 
%E_{(k,\omega(k))} = \frac{\hbar\omega(k)}{8}\left(1 + \frac{k^2}{k_P^2}\right)^2 
%\equiv E_k = \frac{\hbar^2}{8m}\left(k+\frac{k_P^2}{k}\right)^2 \, . 
%\end{eqnarray} 
This again implies the existence of a \emph{pair} of operators:
\begin{eqnarray} \label{U_k_2*}
\hat{U}^{\pm}_{k}(t) = \exp\left[{-i\omega_{\pm}(k)t}\right]\mathbb{I} = \exp\left[- \frac{2 i E_k}{\hbar}\left(1 \pm \sqrt{1 - \frac{\hbar^2\omega_P^2}{4E_k^2}} \right)t\right]\mathbb{I} \, .
\end{eqnarray}
However, as with (\ref{U_k_1}), we may ignore the positive branch as this corresponds to the sub-Planckian regime for black holes, which is governed by Eq.~(\ref{H:m>m_P:w>w_p:k>k_P}). We then have $\hat{U}^{-}_{k}(t) $, as given by Eq. (\ref{U_k_2*}), as the sole unitary time evolution operator for an eigenstate $\phi(x,0) = e^{ikx}$ governed by Eq.~(\ref{H:m>m_P:w<w_p:k<k_P}). In this case, the region of integration for the wave packet (\ref{wavepacket_ext}) is $\Sigma = [-k_{-}(m),k_{-}(m)]$, where $k_{-}(m)$ is given by the dual of Eq. (\ref{disp_rel_m<m_P4}) under the interchange $m \rightarrow m_P'^2/m$. 

By the same reasoning as before, it follows that the time evolution operator for the wave packet $\psi(x,0) = \int_\Sigma a(k)e^{ikx}dk$ is
\begin{eqnarray} \label{U_2}
\hat{U}^{-}_{m>m_P'}(x,t) = \exp\left[- \frac{2 i \hat{H}_{-}'(x)}{\hbar}\left(1 - \sqrt{1 - \frac{\hbar^2\omega_P^2}{4\hat{H}_{-}'^2(x)}} \right)t\right]\mathbb{I} \, ,
\end{eqnarray}
where $\hat{H}_{-}'(x)$ denotes the Hamiltonian which is valid for the negative solution branch, $\omega_{-}(k)$, for $m > m_P'$.  (The prime indicates a regime in which the Hamiltonian is redefined by rewriting the modified Schr{\" o}dinger equation, c.f. Sec. \ref{Sec.3.3}.)  

In the regime containing modes with sub-Planckian wavelengths for $m < m_P'$, we begin by rewriting Eq. (\ref{H:m<m_P:w>w_p:k>k_P}) as
\begin{eqnarray} \label{H:m<m_P:w>w_p:k>k_P-Appendix}
2i\hbar[1- k_P^2/(\partial^2/\partial x^2)]^{-2}\frac{\partial \psi}{\partial t} = \hat{H}'(x,t)\psi \, 
\end{eqnarray}
where
\begin{eqnarray} \label{H:m<m_P:w>w_p:k>k_P-H(x)}
\hat{H}'(x,t) = -4mc^2k_P^{-2}[1- \omega_P^2/(\partial^2/\partial t^2)]^{-1}\frac{\partial^2 }{\partial x^2} \, 
\end{eqnarray}
and the terms in square brackets are defined for $\omega > \omega_P$ and $k > k_P$ via Eqs. (\ref{Exp2A*})-(\ref{Exp2B*}). This implies that the time evolution operator for a single momentum eigenstate $\phi(x,0) = e^{ikx}$ is
\begin{eqnarray} \label{U_k_2**}
\hat{U}_{(k,\omega(k))}(t) = \exp\left[{-i\omega(k)t}\right]\mathbb{I} = \exp\left[- \frac{i E_{(k,\omega(k))}}{2\hbar}\left(1 + \frac{k_P^2}{k^2}\right)^2t\right]\mathbb{I} \, ,
\end{eqnarray}
where $E_{(k,\omega(k))} =  4mc^2(k^2/k_P^2)(1 + \omega_P^2/\omega^2)^{-1} = 2\hbar\omega(k)(1 + k^2/k_P^2)^{-2}$, if we invoke the definition of the Hamiltonian given by Eq.~(\ref{H:m<m_P:w>w_p:k>k_P-H(x)}).
From this, we obtain
\begin{eqnarray} \label{omega(k)_2}
\omega_{\pm}(k) = \frac{\hbar\omega_P^2}{2E_k}\left(1 \pm \sqrt{1 - \frac{4E_k^2}{\hbar^2\omega_P^2}} \right) \, 
\end{eqnarray}
with
\begin{eqnarray} 
E_{(k,\omega(k))} =  2\hbar\omega(k)\left(1 + \frac{k^2}{k_P^2}\right)^{-2}  \equiv E_k = \frac{\hbar^2k_P^4}{2m}\left(k+\frac{k_P^2}{k}\right)^{-2}\, , 
\end{eqnarray} 
which again implies the existence of a \emph{pair} of operators $\hat{U}^{\pm}_{k}(t)$, corresponding to $\hat{U}_{(k,\omega(k))}(t)$. These are identical in form to those given in Eq. (\ref{U_k_1}) but the expression for $E_k$ is different. In this case, the negative branch may be ignored and the same reasoning as before implies that the time evolution operator corresponding to Eq. (\ref{H:m<m_P:w>w_p:k>k_P}) is
\begin{eqnarray} \label{U_3}
\hat{U}^{+}_{m<m_P'}(x,t) = \exp\left[- \frac{i \hbar\omega_P^2}{2\hat{H}_{+}'(x)}\left(1 + \sqrt{1 - \frac{4\hat{H}_{+}'(x)^2}{\hbar^2\omega_P^2}} \right) t\right]\mathbb{I},
\end{eqnarray}
where $\hat{H}_{+}'(x)$ denotes the Hamiltonian for the positive solution branch, $\omega_{+}(k)$, for $m < m_P'$. 

Finally, in the regime containing modes with sub-Planckian wavelengths for $m > m_P'$, we rewrite Eq. (\ref{H:m>m_P:w>w_p:k>k_P}) as
\begin{eqnarray}\label{H:m>m_P:w>w_p:k>k_P-Appendix} 
\frac{i\hbar}{4} \left[1 - \frac{\omega_P^{2}}{ \partial^2 /\partial t^2} \right]\frac{\partial \psi}{\partial t} = \hat{H}(x)\psi \, ,
\end{eqnarray}
where $\hat{H}(x)$ applies to modes with $k > k_P$ via the expansion (\ref{Exp2A*}). This implies the existence of a pair of time evolution operators for single momentum eigenstates $\hat{U}^{\pm}_{k}(t)$ that are identical to those in Eq.~(\ref{U_k_1}) but with 
\begin{eqnarray}
E_k = \frac{\hbar^2}{32m}\left(k + \frac{k_P^2}{k}\right)^2\, .
\end{eqnarray}
It follows that the time evolution operator corresponding to Eq.~(\ref{H:m>m_P:w>w_p:k>k_P}) is
\begin{eqnarray} \label{U_4}
\hat{U}^{+}_{m>m_P'}(x,t) = \exp\left[- \frac{2 i \hat{H}_{+}(x)}{\hbar}\left(1 + \sqrt{1 - \frac{\hbar^2\omega_P^2}{4\hat{H}_{+}^2(x)}} \right) t\right]\mathbb{I} \, ,
\end{eqnarray}
where $\hat{H}_{+}(x)$ denotes the Hamiltonian for the positive solution branch, $\omega_{+}(k)$, for $m > m_P'$. This is simply the Hamiltonian appearing in Eq.~(\ref{H:m>m_P:w>w_p:k>k_P-Appendix}), which automatically applies to modes with $k > k_P$ via the expansion (\ref{Exp2A*}), as stated above. Since no redefinition of the Hamiltonian occurred at the level of the modified Sch{\" o}dinger equation, we do not denote this using a prime. 

Thus there is complete symmetry between the time evolution operators in four different regimes in the sense that
\begin{eqnarray} \label{symmetries-1}
\hat{U}^{\pm}_{m \lesseqgtr m_P'}(x,t) \leftrightarrow \hat{U}^{\pm}_{m \gtreqless m_P'}(x,t) \iff \frac{4\hat{H}^2_{\pm}(x)}{\hbar^2\omega_P^2} 
\leftrightarrow \frac{\hbar^2\omega_P^2}{4\hat{H}'^2_{\pm}(x)}
\end{eqnarray}
and
\begin{eqnarray} \label{symmetries-1}
\hat{U}^{\pm}_{m \lesseqgtr m_P'}(x,t) \leftrightarrow \hat{U}^{\mp}_{m \lesseqgtr  m_P'}(x,t) \iff \hat{H}^2_{\pm}(x) \leftrightarrow \hat{H}'^2_{\mp}(x) \, .
\end{eqnarray}
In the limit 
\begin{eqnarray} \label{limit-1}
m \rightarrow m_P' \iff \omega \rightarrow \omega_P, \ k \rightarrow k_P \ \ \forall \omega, k \, ,
\end{eqnarray}
we have $k_{+}(m_P') = k_{-}(m_P') = k_P$, so that
\begin{eqnarray} \label{limit-2}
\psi(x,0) \rightarrow \phi(x,0) = e^{ik_Px} \, .
\end{eqnarray}
This is the \emph{unique} momentum eigenstate for a particle with mass $m = m_P'$ in the extended de Broglie theory. In this limit, all four Hamiltonians and all four unitary time evolution operators converge, giving
\begin{eqnarray} \label{limit-3}
\hat{H}_{-}(x) = \hat{H}_{-}'(x,t) = \hat{H}_{+}'(x,t) = \hat{H}_{+}(x) = 2m_P'c^2
\end{eqnarray}
and
\begin{eqnarray} \label{limit-4}
\hat{U}^{-}_{m<m_P'}(x,t) = \hat{U}^{-}_{m>m_P'}(x,t) = \hat{U}^{+}_{m<m_P'}(x,t) = \hat{U}^{+}_{m>m_P'}(x,t) = e^{-i\omega_P}.
\end{eqnarray}
%
%%%%%%%%%%%%%%%%%%%%%%%%%%%%%%%%%%%%%%%%%%%%%%%%%%
%%%%%%%%%%%%%%%%%%%%%%%%%%%%%%%%%%%%%%%%%%%%%%%%%%``
\renewcommand{\theequation}{B-\arabic{equation}}
\setcounter{equation}{0}  % reset counter 
\section*{B \ Comparison of the extended de Broglie theory with non-\indent unitary theories containing higher order time derivatives} \label{Appendix B}
The quantum EOM of the extended de Broglie theory involve higher order time derivatives that are often problematic for physical theories requiring both causality \emph{and} unitarity. In particular, such equations appear in various contexts in the literature on quantum gravity, where they often give rise to non-unitary time evolution, at least in a relativistic context. It is therefore instructive to compare these theories with the one considered in this paper, which appears to be unitary in a non-relativistic context.

One example of non-unitary evolution arising from higher order time derivatives occurs in quantum field theories (QFTs) formulated in non-commutative geometry (NCG). In this scenario, the break down of unitarity at the perturbative level is closely tied to the violation of causality. Though space-space non-commutativity may be consistent with causality, even if local Lorentz invariance is broken, time-space non-commutativity is problematic, implying the existence of particles that are effectively extended in time \cite{AlvarezGaume:2001ka,AlvarezGaume:2000bv}. Heuristically, non-commutative particles behave like rigid extended dipoles, oriented in the direction of the four-vector $L^{\mu} = \theta^{\mu\nu}p_{\nu}$, where $p^{\mu}$ is the four-momentum of the state and $\theta^{\mu\nu} \propto [x^{\mu},x^{\nu}]$ is the position coordinate commutator \cite{Seiberg:2000gc}. Intuitively, it is straightforward to see how such non-locality in time may lead to advanced effects, in which events precede their apparent causes. For example, suddenly displacing one end of a space-like rod, the signal instantaneously appears at the opposite end \cite{AlvarezGaume:2000bv}. Thus, for a non-commutative particle reflected by a potential barrier, the centre of mass then recoils \emph{before} the particle hits the wall \cite{Seiberg:2000gc}. 

Technically, such theories have no canonical Hamiltonian quantization and must be defined operationally via Feynman diagram expansions. Unitary evolution is therefore not assured, and must be checked explicitly (see, for example, \cite{Bahns:2002vm} and references therein). Obtaining the propagators by summation at tree level, various important physical properties of a given theory can be determined, such as the effective Lagrangian and  dispersion relations, which must be satisfied by the EOM \cite{AlvarezGaume:2000bv}. In \cite{AlvarezGaume:2001ka}, it was explicitly shown that propagators in theories with time-space non-commutativity imply `sick' dispersion relations, giving rise to tachyonic states which automatically violate unitarity. The same relations imply EOM with higher order time derivatives, so that violations of causality/unitarity can be seen as resulting from the corresponding non-locality in time. Specifically, if the action is \emph{arbitrarily} non-local in time, the evolution at of the fields at time $t'$ depends on their configurations at both $t < t'$ and $t > t'$.

From this brief discussion, it is clear that several technical and conceptual issues are inherently linked. The Hamiltonian and path integral quantization of classical fields are inequivalent in geometries with time-space non-commutativity. Hamiltonian quantization is is not well defined, whereas operational definitions via Feynman diagram expansions lead to unphysical dispersion relations. Specifically, these imply EOM with higher order time derivates which result in violations of both causality and unitarity, manifested in the existence of tachyonic particles, or states with negative norm. However, this discussion also highlights the fact that non-unitarity may occur in \emph{any} theory containing higher order derivatives in time/modified dispersion relations. The question then arises of why the theory of extended de Broglie relations avoids these problems. 

Two possibilities suggest themselves. The first is that the extended theory escapes such difficulties simply because it is formulated in a non-relativistic context. As stated in \cite{Seiberg:2000gc}, violations of causality are not disastrous for non-relativistic field theories. This is also true for Newtonian gravity, even though it implies instantaneous action at a distance. If, in general, acausal behavior and non-unitarity arise from the same structural modifications of canonical QFT, namely modified dispersion relations and the resulting non-locality in time (whether or not this occurs due to NCG), then selecting only leading order terms in the expansion for $\omega(k)$, corresponding to the Newtonian approximation, may neglect terms leading to the violation of unitarity.  In this case, we would expect to find a consistent Hamiltonian quantization, as given in Sec. \ref{Sec.3.3}. If unitarity is maintained only to leading order, attempts to generalize the present work to include relativistic effects may lead to inconstancies similar to those in NCG.

The second (more interesting) possibility is that the extend de Broglie theory, as formulated non-relativistically, \emph{already} incorporates certain effects associated with the correct relativistic theory, which ensure the preservation of unitarity. The right-hand side of the $(k,\omega(k))$ diagram corresponds to the quantization of \emph{extended} objects and manifestly mixes IR/UV effects due to the inclusion of observable quantities of the form $O' \sim O^{-1} + O$. By contrast, canonical quantization of the left-hand side corresponds to the quantization of point particles, so that the extended de Broglie theory effectively treats \emph{all} objects as inherently extended in \emph{space} at the classical level. While IR/UV mixing is typical of \emph{all} theories in NCG, the mathematical and structural foundations of those found to violate unitarity/causality are otherwise based on the quantization of point-like particles. 

It is interesting that the other known examples of inherently non-commutative backgrounds occur in the theory of quantized open strings, which do \emph{not} reduce to canonical QFTs in the IR limit. In other words, the effects of stringy excitations cannot be neglected, even at low energies. In addition, such excitations are key to the preservation of unitarity, conspiring to exactly cancel the acausal effects that are present for canonical fields in NCG \cite{Seiberg:2000gc}. Intuitively, this can be explained as follows. When two strings scatter off one another, an intermediate stretched string is formed. If this has total energy $E$, its length $L(t)$ oscillates between small and large values such that $E \sim L/l_s^2 + N/L$, where $l_s^2$ is the fundamental string length scale and $N$ is the oscillation number \cite{Seiberg:2000gc}. The string expands to a maximal size $L \sim El_s^2$, storing the energy as potential, before the oscillation repeats. In each oscillation there is a finite probability for the string to split, resulting in an infinite sequence of delayed wave packets, separated by time interval obeying the stringy uncertainty relation \cite{Seiberg:2000gc}. This exactly cancels the advanced effects induced by the time non-locality of the NCG. 

Such behavior is possible only for an extended object, and it is intriguing that the energy and length of the intermediate string state obeys the `half' T-duality symmetry characteristic of the extended de Broglie relations proposed in the present work. Indeed, these proposals were initially motivated by the desire to mix IR and UV effects by providing a unified framework in which to consider both black holes and fundamental particles. As we have seen, this is necessarily equivalent to considering `particles' as being inherently extended in some way, prior to quantization.  

Although further work is required to establish whether this is indeed the case, we may conjecture that the form of the extended de Broglie relations n{\" a}ively captures certain \emph{irremovable} features of the quantization of extended objects, which manifest even in the IR theory. Optimistically, such relations may even help alleviate some of the theoretical difficulties associated with the quantization of macroscopic objects in general (c.f. \cite{Feynman:1996kb}).   

%
%%%%%%%%%%%%%%%%%%%%%%%%%%%%%%%%%%%%%%%%%%%%%%%%%%
%%%%%%%%%%%%%%%%%%%%%%%%%%%%%%%%%%%%%%%%%%%%%%%%%%``
\begin{center}
{\bf Acknowledgments}
\end{center}
We thank the Research Center for the Early Universe (RESCEU), University of Tokyo, for gracious hospitality during the preparation of this manuscript. M.L. wishes to thank Tiberiu Harko for interesting and helpful discussions.
%
%%%%%%%%%%%%%%%%%%%%%%%%%%%%%%%%%%%%%%%%%%%%%%%%%%
%%%%%%%%%%%%%%%%%%%%%%%%%%%%%%%%%%%%%%%%%%%%%%%%%%``

 %

\begin{thebibliography}{99}

%\cite{Rae00}
\bibitem{Rae00}
A.~I.~M.~Rae, 
{\it Quantum Mechanics, $4^{th}$ Ed.},
Institute of Physics Publishing (2000).

%\cite{Ish95}
\bibitem{Ish95}
C.~J.~Isham, 
{\it Lectures on Quantum Theory: Mathematical and Structural Foundations},
Imperial College Press (1995).
% \cite{Rae00,Ish95}

%\cite{Ca:2014}
\bibitem{Ca:2013}
B. J. Carr, 
{\it Black Holes, the Generalized Uncertainty Principle and Higher Dimensions},
Mod. Phys. Lett. {\bf A 28}, 1340011 (2013).

%\cite{Ca:2014}
\bibitem{Ca:2014}
B.~J.~Carr, {\it The Black Hole Uncertainty Principle correspondence}, In 1st Karl Schwarzschild meeting on gravitational physics, 
ed. P.~Nicolini, M.~Kaminski, J.~Mureilka , M. Bleicher, pp. 159-167 (Springer).

%\cite{Ca:2014}
%\bibitem{Ca:2014}
%B.~Carr,  
%{\it The Black Hole Uncertainty Principle Correspondence}, 
%Proceedings of Schwarzschild meeting (Frankfurt July 2013), arXiv:1402.1427  (2014).

%\cite{cmp}
\bibitem{cmp}
B. J. Carr, L. Modesto and I. Pr\'emont-Schwarz, 
{\it Generalized Uncertainty Principle and self-dual black holes}, 
arXiv: 1107.0708 (2011). 

%\cite{Zurek:2003zz}
\bibitem{Zurek:2003zz} 
W.~H.~Zurek,
{\it Decoherence, einselection, and the quantum origins of the classical},
Rev.\ Mod.\ Phys.\  {\bf 75}, 715 (2003).

%\cite{Anglin:1996bb}
\bibitem{Anglin:1996bb} 
J.~R.~Anglin, J.~P.~Paz and W.~H.~Zurek,
{\it Deconstructing decoherence},
Phys.\ Rev.\ A {\bf 55}, 4041 (1997),
[arXiv:quant-ph/9611045].
%\cite{Zurek:2003zz,Anglin:1996bb}

%\cite{Singh:2015sua}
\bibitem{Singh:2015sua} 
T.~P.~Singh,
{\it Possible role of gravity in collapse of the wave-function: a brief survey of some ideas},
arXiv:1503.01040 [quant-ph].

%\cite{Tawfik:2014zca}
\bibitem{Tawfik:2014zca} 
A.~N.~Tawfik and A.~M.~Diab,
{\it Generalized Uncertainty Principle: Approaches and Applications},
Int.\ J.\ Mod.\ Phys.\ D {\bf 23}, 1430025 (2014)
[arXiv:1410.0206 [gr-qc]].

%\cite{LaCa:2015}
\bibitem{LaCa:2015}
M.~J.~Lake and B.~Carr,  
{\it The black hole uncertainty principle  correspondence in higher dimensions}, in preparation (2015).

%\cite{La15}
\bibitem{La15}
M.~J.~Lake,
{\it Instantaneous measurements of nonlocal variables in relativistic quantum theory (a review)}, 
arXiv:1505.05052 [quant-ph].
%\cite{LaCa:2015,La15}

%\cite{Pe11}
\bibitem{Pe11}
P.~Pedram, 
{\it A class of GUP solutions in deformed quantum mechanics},
Int. J. Mod. Phys. D 19:2003-2009 (2010),
arXiv:1103.3805 [hep-th].

%\cite{La09}
\bibitem{La09}
A.~Lavagno, 
{\it Basic-deformed quantum mechanics},
Rept. Math. Phys. 64, 1-2, 79-91(2009),
arXiv:1103.3805 [hep-th].

%\cite{Zhang:2003wv}
\bibitem{Zhang:2003wv} 
J.-Z.~Zhang,
{\it A $q$-deformed quantum mechanics},
Phys.\ Lett.\ B {\bf 440}, 66 (1998)
[hep-th/0310043].

%\cite{Hirshfeld(2002)}
\bibitem{Hirshfeld(2002)} A.~C.~Hirshfeld and P.~ Henselder, 
{\it Deformation quantization in the teaching of quantum mechanics}
American Journal of Physics, {\bf 70}, 537 (2002).
%\cite{La09,Zhang:2003wv,Hirshfeld(2002)}

%\cite{Maziashvili:2011dx}
\bibitem{Maziashvili:2011dx} 
M.~Maziashvili,
{\it Implications of minimum-length deformed quantum mechanics for QFT/QG},
Fortsch.\ Phys.\  {\bf 61}, 685 (2013)
[arXiv:1110.0649 [gr-qc]].

%\cite{Faizal:2015kqa}
\bibitem{Faizal:2015kqa} 
M.~Faizal,
{\it Deformation of Second and Third Quantization},
arXiv:1503.04797 [gr-qc].

%\cite{Soviet}
\bibitem{Soviet}
V.~I.~Grigoriev,
in {\it The Great Soviet Encyclopedia}, 3rd Edition (1970-1979).

%\cite{Greiner:1990tz}
\bibitem{Greiner:1990tz} 
W.~Greiner,
{\it Relativistic quantum mechanics: Wave equations},
Berlin, Germany: Springer (Theoretical physics, 3)  (1990).

%\cite{AlvarezGaume:2012zz}
\bibitem{AlvarezGaume:2012zz} 
L.~Alvarez-Gaume and M.~A.~Vazquez-Mozo,
{\it An invitation to quantum field theory},
Lect.\ Notes Phys.\  {\bf 839}, 1 (2012).
%\cite{Soviet,{Greiner:1990tz,AlvarezGaume:2012zz}

%\cite{Padmanabhan:1986ny}
\bibitem{Padmanabhan:1986ny} 
T.~Padmanabhan,
{\it Physical Significance of Planck Length},
Annals Phys.\  {\bf 165}, 38 (1985).

%\cite{Hinojosa:2015tga}
\bibitem{Hinojosa:2015tga} 
C.~B.~Hinojosa and J.~L{\' o}pez-Sarri{\' o}n,
{\it Moving Schwarzschild Black Hole and Modified Dispersion Relations},
arXiv:1503.05593 [gr-qc].

%\cite{Laperashvili:2015pea}
\bibitem{Laperashvili:2015pea} 
L.~V.~Laperashvili, H.~B.~Nielsen and B.~G.~Sidharth,
{\it Planck Scale Physics, Gravi-Weak Unification and the Higgs Inflation},
arXiv:1503.03911 [gr-qc].

\bibitem{cmn}
B.~J.~Carr, J.~Mureika and P.~Nicolini, 
{\it Sub-Planckian black holes and the Generalized Uncertainty Principle},
[arXiv:1504.07637] (2015).

%\bibitem{ling1}
%Y.~Ling, B.~Hu and X.~Li,
%{\it Modified dispersion relations and black hole physics},
%Phys.\ Rev.\ D {\bf 73}, 087702 (2006).

%\bibitem{ling2}
%X.~Han, H.~r.~Li and Y.~Ling,
%{\it Modified dispersion relations and (A)dS Schwarzschild Black holes},
%Phys.\ Lett.\ B {\bf 666}, 121 (2008).
%[arXiv:0807.4269 [gr-qc]].

%\bibitem{ling3}
%L.~Xiang, Y.~Ling and Y.~G.~Shen,
%{\it Singularities and the Finale of Black Hole Evaporation},
%Int.\ J.\ Mod.\ Phys.\ D {\bf 22}, 1342016 (2013).
%[arXiv:1305.3851 [gr-qc]].

%\cite{Chang:2011jj}
\bibitem{Chang:2011jj} 
  L.~N.~Chang, Z.~Lewis, D.~Minic and T.~Takeuchi,
  {\it On the Minimal Length Uncertainty Relation and the Foundations of String Theory},
  Adv.\ High Energy Phys.\  {\bf 2011}, 493514 (2011)
  [arXiv:1106.0068 [hep-th]].
  
%\cite{Benczik:2002tt}
\bibitem{Benczik:2002tt} 
  S.~Benczik, L.~N.~Chang, D.~Minic, N.~Okamura, S.~Rayyan and T.~Takeuchi,
  {\it Short distance versus long distance physics: The Classical limit of the minimal length uncertainty relation},
  Phys.\ Rev.\ D {\bf 66}, 026003 (2002)
  [hep-th/0204049].
  
%\cite{Benczik:2002px}
\bibitem{Benczik:2002px} 
  S.~Benczik, L.~N.~Chang, D.~Minic, N.~Okamura, S.~Rayyan and T.~Takeuchi,
  {\it Classical implications of the minimal length uncertainty relation},
  hep-th/0209119.
%\cite{Chang:2011jj,Benczik:2002tt,Benczik:2002px}

%\cite{Garay:1994en}
\bibitem{Garay:1994en} 
  L.~J.~Garay,
  {\it Quantum gravity and minimum length},
  Int.\ J.\ Mod.\ Phys.\ A {\bf 10}, 145 (1995)
  [gr-qc/9403008].
  
%\cite{Hossenfelder:2012jw}
\bibitem{Hossenfelder:2012jw} 
  S.~Hossenfelder,
  {\it Minimal Length Scale Scenarios for Quantum Gravity},
  Living Rev.\ Rel.\  {\bf 16}, 2 (2013)
  [arXiv:1203.6191 [gr-qc]]. 
%\cite{Garay:1994en,Hossenfelder:2012jw}

%\cite{Raghavan:2012sy}
\bibitem{Raghavan:2012sy} 
  R.~S.~Raghavan, D.~Minic, T.~Takeuchi and C.~H.~Tze,
  {\it Using Neutrinos to test the Time-Energy Uncertainty Relation in an Extreme Regime},
  arXiv:1210.5639 [hep-ph].
  
%\cite{AmelinoCamelia:1997em}
\bibitem{AmelinoCamelia:1997em} 
  G.~Amelino-Camelia,
  {\it Classicality, matter - antimatter asymmetry, and quantum gravity deformed uncertainty relations},
  Mod.\ Phys.\ Lett.\ A {\bf 12}, 1387 (1997)
  [gr-qc/9706007].

%\cite{Adler_1} 
\bibitem{Adler_1} 
R.~J.~Adler and D.~I.~Santiago, 
{\it On gravity and the Uncertainty Principle}, 
Mod. Phys. Lett. {\bf A14}, 1371 (1999).

%\cite{Adler_2} 
\bibitem{Adler_2}
R.~J.~Adler, P.~Chen and D.~I.~Santiago, 
{\it The Generalized Uncertainty Principle and black hole remnants}, 
Gen. Rel. Grav. {\bf 33}, 2101 (2001).

%\cite{Adler_3} 
\bibitem{Adler_3} 
P.~Chen and R.~J.~Adler, 
{\it Black hole remnants and dark matter},  
Nucl.Phys.Proc.Suppl. \textbf{124}  103 (2003). 

%\cite{Adler_4} 
\bibitem{Adler_4}
R.~J.~Adler, 
{\it Six easy roots to the Planck scale},
Am.\ J.\ Phys.\  {\bf 78}, 925 (2010).
  
%\cite{Ro29}
\bibitem{Ro29}
H.~P.~Robertson, 
{\it The Uncertainty Principle},
Phys. Rev. 34, 163-64 (1929).

%\cite{Sc30}
\bibitem{Sc30}
E.~Schr{\" o}dinger, 
{\it Zum Heisenbergschen Unsch�rfeprinzip},
Phys. Math. Klass. 14, 296-303 (1930).
%\cite{Ro29,Sc30}

%\cite{Oz03A}
\bibitem{Oz03A}
M.~Ozawa, 
{\it Physical content of Heisenberg's uncertainty relation: Limitation and reformulation},
Phys. Lett. A 318, 21-29 (2003),
[arXiv:quant-ph/0210044].

%\cite{Oz03B}
\bibitem{Oz03B}
M.~Ozawa, 
{\it Universally valid reformulation of the Heisenberg uncertainty principle on noise and disturbance in measurement},
Phys. Rev. A 67, 042105, (1-6) (2003),
[arXiv:quant-ph/0207121].
%\cite{Oz03A,Oz03B}

%\cite{ashtekar_1}
\bibitem{ashtekar_1}
A.~Ashtekar, S.~Fiarhurst and J.~L.~Willis, 
{\it Quantum gravity, shadow states and quantum mechanics}, 
Class. Quant. Grav. {\bf 20}, 1031 (2003).
%G. M. Hossain, V. Husain and S. S. Seahra, Class. Quant. Grav. {\bf 207}, 165013 (2010).

%\cite{ashtekar_2}
\bibitem{ashtekar_2}
%A. Ashtekar, S. Fiarhurst and J. L. Willis, Class. Quant. Grav. {\bf 20}, 1031 (2003);
G.~M.~Hossain, V.~Husain and S.~S.~Seahra, 
{\it Background independent quantization and the uncertainty principle}, 
Class. Quant. Grav. {\bf 207}, 165013 (2010).

%\cite{veneziano_1}
\bibitem{veneziano_1}
G.~Veneziano, 
{\it A stringy nature needs just two constants}, 
Europhys. Lett. {\bf 2}, 199 (1986). 

%\cite{veneziano_2}
\bibitem{veneziano_2} 
E.~Witten, 
{\it Reflections on the fate of space-time}, 
Phys.~Today {\bf 49N4}, 24-30 (1996).

%\cite{veneziano_3}
\bibitem{veneziano_3}
F.~Scardigli, 
{\it Generalized uncertainty principle in quantum gravity from micro-black hole Gedanken experiment}, 
Phys.~Lett.~ {\bf B452}, 39 (1999).

%\cite{veneziano_4}
\bibitem{veneziano_4}
D.~J.~Gross and P.~F.~Mende, 
{\it String theory beyond the Planck scale}, 
Nuc. Phys. {\bf B303}, 407 (1988).

%\cite{veneziano_5}
\bibitem{veneziano_5}
D.~Amati, M.~Ciafaloni and G.~Veneziano, 
{\it Can spacetime be probed below the string size?}, 
Phys, Lett. {\bf B216}, 41 (1989).

%\cite{veneziano_6}
\bibitem{veneziano_6}
 T.~Yoneya, 
 {\it On the interpretation of minimal length in string theories}, 
 Mod. Phys. Lett. {\bf A4}, 1587 (1989)

%\cite{majid}
\bibitem{majid} 
S.~Majid, 
{\it Scaling limit of the non-commutative black hole}, 
J. Phys. Conf. Ser. 284, 012003 (2011).

%\cite{nicolini}
\bibitem{nicolini} 
M.~Isi, J.~Mureika and P.~Nicolini, 
{\it Self-completeness and the generalized uncertainty}, 
JHEP~{\bf 1311}, 139 (2013).

%\cite{maggiore_1}
\bibitem{maggiore_1}
M.~Maggiore, 
{\it A generalized uncertainty principle in quantum gravity}, 
Phys. Lett. {\bf B 304}, 65 (1993).

%\cite{maggiore_2}
\bibitem{maggiore_2}
M.~Maggiore, 
{\it The algebraic structure of the generalized uncertainty principle}, 
Phys. Lett. {\bf B 319}, 83 (1993).

%\cite{maggiore_3}
\bibitem{maggiore_3}
M.~Maggiore, 
{\it Quantum groups, gravity and the generalized uncertainty principle}, 
Phys. Rev. {\bf D 49}, 5182 (1994).

%\cite{AmelinoCamelia:2002wr}
\bibitem{AmelinoCamelia:2002wr} 
  G.~Amelino-Camelia,
  {\it Doubly special relativity},
  Nature {\bf 418}, 34 (2002)
  [gr-qc/0207049].
  
%\cite{rainbow}   
\bibitem{rainbow} 
J.~Magueijo and L.~Smolin, 
{\it Gravity's Rainbow},
Class.~Quant.~Grav.~{\bf 21}, 1725-1736 (2004). 
%[arXiv:gr-qc/0305055].

%%%%%%%%%%%%%%%%%%%References for Appendix B

%\cite{AlvarezGaume:2001ka}
\bibitem{AlvarezGaume:2001ka} 
  L.~Alvarez-Gaume, J.~L.~F.~Barbon and R.~Zwicky,
  {\it Remarks on time space noncommutative field theories},'
  JHEP {\bf 0105}, 057 (2001)
  [hep-th/0103069].
  
%\cite{AlvarezGaume:2000bv}
\bibitem{AlvarezGaume:2000bv} 
  L.~Alvarez-Gaume and J.~L.~F.~Barbon,
  {\it Nonlinear vacuum phenomena in noncommutative QED},
  Int.\ J.\ Mod.\ Phys.\ A {\bf 16}, 1123 (2001)
  [hep-th/0006209].
  
%\cite{Seiberg:2000gc}
\bibitem{Seiberg:2000gc} 
  N.~Seiberg, L.~Susskind and N.~Toumbas,
  {\it Space-time noncommutativity and causality},
  JHEP {\bf 0006}, 044 (2000)
  [hep-th/0005015].
  
%\cite{Bahns:2002vm}
\bibitem{Bahns:2002vm} 
  D.~Bahns, S.~Doplicher, K.~Fredenhagen and G.~Piacitelli,
  {\it On the Unitarity problem in space-time noncommutative theories},
  Phys.\ Lett.\ B {\bf 533}, 178 (2002)
  [hep-th/0201222].
  
%\cite{Feynman:1996kb}
\bibitem{Feynman:1996kb} 
  R.~P.~Feynman, F.~B.~Morinigo, W.~G.~Wagner and B.~Hatfield,
  {\it Feynman lectures on gravitation},
  Reading, USA: Addison-Wesley (1995) 232 p. (The advanced book program)

 \end{thebibliography}
\end{document}